%% file: main.tex
\documentclass{nature}

\usepackage{amsmath}
\usepackage{amssymb}
\usepackage{hyperref}
\usepackage{lineno}
\usepackage{float}

\hypersetup{colorlinks,citecolor=blue,linkcolor=blue,urlcolor=blue}

\DeclareUnicodeCharacter{2212}{-}

\title{A long-period radio transient active for three decades}

\author{N. Hurley-Walker$^{*1}$,
N.~Rea$^{2,3}$, 
S.~J.~McSweeney$^{1}$, 
B.~W.~Meyers$^{1}$, 
E.~Lenc$^{4}$, 
I.~Heywood$^{5,6,7}$, 
S.~D.~Hyman$^8$, 
Y.~P.~Men$^9$, 
T.~E.~Clarke$^{10}$, 
F.~Coti Zelati$^{2,3}$, 
D.~C.~Price$^{1}$, 
C.~Horv\'{a}th$^{1}$, 
T.~J.~Galvin$^{11, 1}$, 
G.~E.~Anderson$^{1}$, 
A.~Bahramian$^{1}$, 
E.~D.~Barr$^9$, 
N.~D.~R.~Bhat$^{1}$, 
M.~Caleb$^{12,13}$,
M. Dall'Ora$^{14}$, 
D.~de Martino$^{14}$, 
S.~Giacintucci$^{10}$, 
J.~S.~Morgan$^{1}$, 
K.M.~Rajwade$^{15}$,
B.~Stappers$^{16}$,
A.~Williams$^{1}$ 
}

\usepackage{graphicx}
\makeatletter
\let\saved@includegraphics\includegraphics
\AtBeginDocument{\let\includegraphics\saved@includegraphics}
\renewenvironment*{figure}{\@float{figure}}{\end@float}
\makeatother

\begin{document}

\newcommand{\fig}{Fig.}
\newcommand{\figs}{Figs.}
\newcommand{\Fig}{Fig.}
\newcommand{\EFig}{Extended Data Fig.}
\newcommand{\Figs}{Figs.}
\newcommand{\sect}{Section}
\newcommand{\sects}{Sections}
\newcommand{\Sect}{Section}
\newcommand{\Sects}{Sections}
\newcommand{\tab}{Table}
\newcommand{\tabs}{Tables}
\newcommand{\Tab}{Table}
\newcommand{\Tabs}{Tables}
\newcommand{\eqn}{equation}
\newcommand{\eqns}{equations}
\newcommand{\Eqn}{Equation}
\newcommand{\Eqns}{Equations}
\newcommand{\etal}{et~al.}

\newcommand{\farcm}{\mbox{\ensuremath{.\mkern-4mu^\prime}}}
\newcommand{\farcs}{\mbox{\ensuremath{.\!\!^{\prime\prime}}}}
\newcommand{\fdg}{\mbox{\ensuremath{.\!\!^\circ}}}

\newcommand{\src}{GPM\,J\,1839\ensuremath{-}10}
\newcommand{\GLX}{GLEAM-X\,J\,162759.5\ensuremath{-}523504.3}
\newcommand{\PSR}{PSR\,J\,0901\ensuremath{-}4046}
\newcommand{\srcDM}{\ensuremath{273.5}}
\newcommand{\DMerror}{\ensuremath{2.6}}
\newcommand{\DMunit}{pc\,cm\ensuremath{^{-3}}}
\newcommand{\srcDist}{\ensuremath{5.7}}
\newcommand{\Disterror}{\ensuremath{2.9}}
\newcommand{\Distunit}{kpc}
\newcommand{\srcRM}{\ensuremath{-573}}
\newcommand{\RMerror}{\ensuremath{1}}
\newcommand{\RMunit}{rad\,m\ensuremath{^{-2}}}
\newcommand{\srcPlong}{\ensuremath{1318.1957}}
\newcommand{\srcPerr}{\ensuremath{0.0002}}
\newcommand{\srcP}{\ensuremath{1318}}
\newcommand{\srcFlong}{\ensuremath{0.000758612(7)}}
\newcommand{\srcFerr}{\ensuremath{1.2\times10^{-10}}}
\newcommand{\srcFdotlong}{\ensuremath{5\times10^{-20}}}

\newcommand{\srcalpha}{\ensuremath{-3.17}}
\newcommand{\srcalphaerr}{\ensuremath{0.06}}
\newcommand{\srcq}{\ensuremath{-0.56}}
\newcommand{\srcqerr}{\ensuremath{0.03}}

\newcommand{\ndetections}{71}
\newcommand{\srcPdot}{\ensuremath{3.6\times10^{-13}}}
\newcommand{\srcPdotts}{\ensuremath{9.5\times10^{-13}}}
\newcommand{\Pdotunit}{\,s\,s\ensuremath{^{-1}}}

\newcommand{\srcBp}
{\ensuremath{1.2\times10^{15}}}
\newcommand{\srcBpts}
{\ensuremath{2.4\times10^{15}}}

\newcommand{\srcBpwd}{\ensuremath{4\times10^{9}}}
\newcommand{\srcBptswd}{\ensuremath{8.1\times10^{9}}}

\newcommand{\srcSLum}{\ensuremath{1.3\times10^{25}}}
\newcommand{\srcSLumts}{\ensuremath{6\times10^{25}}}

\newcommand{\srcSLumwd}{\ensuremath{2.4\times10^{30}}}
\newcommand{\srcSLumtswd}{\ensuremath{1\times10^{31}}}

\newcommand{\srctau}{\ensuremath{58}}
\newcommand{\srctauts}{\ensuremath{22}}

\newcommand{\srcRLum}{\ensuremath{10^{28}}}

\newcommand{\ergpers}{erg\,s\ensuremath{^{-1}}}
\newcommand{\flux}{erg\,cm\ensuremath{^{-2}}\,s\ensuremath{^{-1}}}

\def\arc{\mbox{$^{\prime\prime}$}}
\def\nh{\hbox{$N_{\rm H}$}}

\maketitle

\begin{affiliations}
 \item International Centre for Radio Astronomy Research, Curtin University, Kent St, Bentley WA 6102, Australia
  \item Institute of Space Sciences (ICE), CSIC, Campus UAB, Carrer de Can Magrans s/n, E-08193, Barcelona, Spain
 \item Institut d'Estudis Espacials de Catalunya (IEEC), Carrer Gran Capit\`a 2--4, E-08034 Barcelona, Spain
 \item ATNF, CSIRO Space \& Astronomy, PO Box 76, Epping, NSW 1710, Australia
 \item Astrophysics, University of Oxford, Denys Wilkinson Building, Keble Road, Oxford OX1 3RH, UK
 \item Department of Physics and Electronics, Rhodes University, PO Box 94, Makhanda 6140, South Africa
 \item South African Radio Astronomy Observatory (SARAO), 2 Fir Street, Observatory, Cape Town 7925, South Africa
 \item Department of Engineering and Physics, Sweet Briar College, Sweet Briar, VA 24595, USA
  \item Max-Planck-Institut f\"ur Radioastronomie, Auf dem H\"ugel 69, D-53121 Bonn, Germany
  \item Remote Sensing Division, U.S. Naval Research Laboratory, Washington, DC 20375, USA
   \item CSIRO, Space and Astronomy, PO Box 1130, Bentley WA 6102, Australia
  \item Sydney Institute for Astronomy, School of Physics, The University of Sydney, Sydney, 2006 NSW, Australia
   \item ASTRO3D: ARC Centre of Excellence for All-sky Astrophysics in 3D, Canberra, 2601 ACT, Australia
  \item INAF – Capodimonte Astronomical Observatory Naples Via Moiariello 16, I-80131 Naples, Italy
 \item ASTRON, the Netherlands Institute for Radio Astronomy, Oude Hoogeveensedĳk 4, 7991 PD Dwingeloo, The Netherlands
 \item Jodrell Bank Centre for Astrophysics, Department of Physics and Astronomy, The University of Manchester, Manchester M13 9PL, UK

\end{affiliations}

\begin{abstract}

Recently several long-period radio transients have been discovered, with strongly polarised coherent radio pulses appearing on timescales between tens to thousands of seconds \cite{2022NatAs...6..828C,2022Natur.601..526H}. In some cases the radio pulses have been interpreted as coming from rotating neutron stars with extremely strong magnetic fields, known as magnetars; the origin of other, occasionally periodic and less well-sampled radio transients, is still debated\cite{2005Natur.434...50H}. Coherent periodic radio emission is usually explained by rotating dipolar magnetic fields and pair production mechanisms, but such models do not easily predict radio emission from such slowly-rotating neutron stars and maintain it for extended times. On the other hand, highly magnetic isolated white dwarfs would be expected to have long spin periodicities, but periodic coherent radio emission has not yet been directly detected from these sources. Here we report observations of a long-period (21 minutes) radio transient, which we have labeled \src{}. The pulses vary in brightness by two orders of magnitude, last between 30 and 300 seconds, and have quasi-periodic substructure. The observations prompted a search of radio archives, and we found that the source has been repeating since at least 1988. The archival data enabled constraint of the period derivative to $<$\srcPdot{}\,\Pdotunit{}, which is at the very limit of any classical theoretical model that predicts dipolar radio emission from an isolated neutron star.

\end{abstract}

We monitored the Galactic Plane using the Murchison Widefield Array (MWA; see Methods) to detect ``slow'' transients within 48 hours of observation. From this search, we discovered the transient source \src{}, initially with two significant detections of 30-s wide pulses on a single night. As monitoring continued, we made several further detections, and also commenced a continuous monitoring campaign with the MWA, and follow-up observations with the Australia Telescope Compact Array, Parkes/Murriyang Radio Telescope, the Australia Square Kilometer Array Pathfinder, and MeerKAT.

\begin{figure}[H]
\centering
\includegraphics[width=89mm]{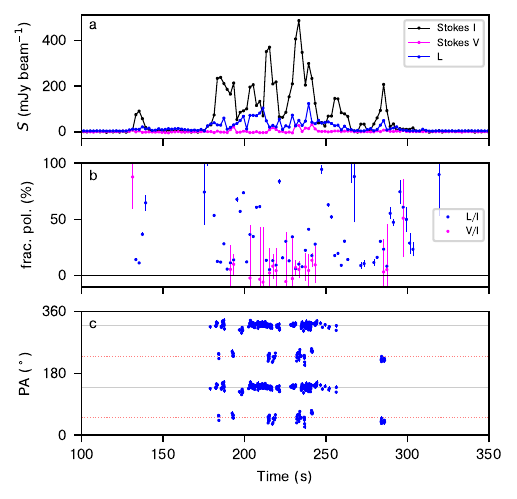}
\caption{Polarisation properties of the brightest pulse observed with MeerKAT, in the observation starting at UTC 2022-07-20~19:12:33. Panel (a) shows the Stokes~I flux density, circular polarised intensity Stokes~V, and linear polarised intensity $P$; panel (b) shows their ratio, i.e. the fractional polarisation; and panel (c) shows the polarisation position angle, derived from the PTUSE data at 65\,ms resolution. For clarity, in the bottom two panels, only data where the errors are less than 100\,\% are shown. In the third panel, the grey line shows the typical (arbitrary) position angle of the polarisation, while the red dotted line shows orthogonal ($\pm$90$^\circ$) angles.\label{fig:meerkat_pol}}
\end{figure}

Interferometric observations with MeerKAT constrained the (J2000) position to RA$=$18$^\mathrm{h}$39$^\mathrm{m}$02.$^\mathrm{s}$00, Dec$=-10^\circ$31$'$49\farcs37, with an uncertainty of 0\farcs15, while the high time resolution (3.9-ms) 
observing mode enabled a detailed view of the pulse structure, and an accurate calculation of the dispersion measure: \srcDM{}$\pm$\DMerror{}\,\DMunit{}. Using the YMW16 Galactic free electron density model\cite{2017ApJ...835...29Y}, this corresponds to a distance of \srcDist{}$\pm$\Disterror{}\,\Distunit{} for the line-of-sight towards \src{}. Over a typical pulse, the degree of linear polarisation varies from 10--100\,\% with a generally flat phase angle.
We observe bursts of duration 0.2--4\,s in which this angle switches by 90$^\circ$, resembling the orthogonal polarization modes (OPMs) commonly seen in neutron star pulsars\cite{Ekers1969} (\Fig~\ref{fig:meerkat_pol}). Quasi-periodic oscillations are seen in many pulses, reminiscent of pulsar single-pulse microstructure, but two to three orders of magnitude larger in scale than expected (see Methods for more details on pulse polarisation and substructure).

\begin{figure}[H]
\centering
\includegraphics[width=89mm]{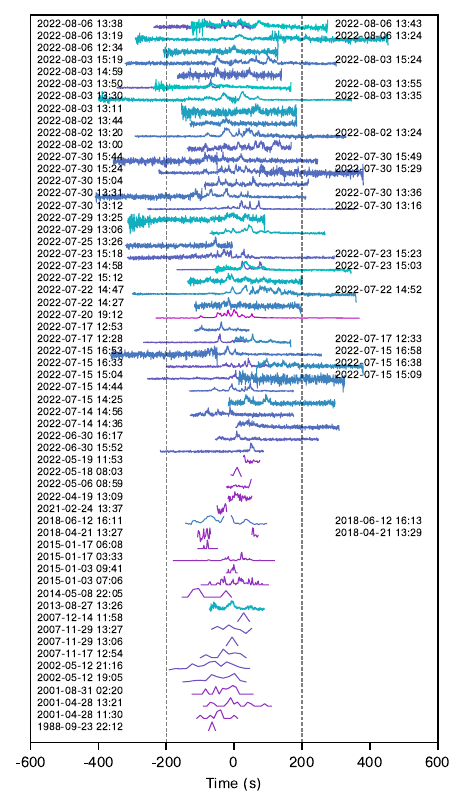}
\caption{78~representative pulses of \src{} aligned according to its measured period $P$ (and $\dot{P}$ set to zero). Flux densities are normalised to the peak of each pulse for readability; barycentric and dispersive corrections have been applied. The observation start times in UTC are listed on the left or right of each detection. The observations were drawn from many observing programmes in various archives and therefore some pulses are not fully sampled. Dashed grey vertical lines indicate the $\sim$400-s wide window in which pulses have been observed to appear. The color range spans 88\,MHz (cyan) to 500\,MHz (magenta) and the detections span 33.9\,years. \label{fig:light_curves}}
\end{figure}

Searches of archival data yielded further detections as far back as 1988 by the Very Large Array (VLA), the VLA Low-band Ionosphere and Transient Experiment (VLITE), and the Giant Metrewave Radio Telescope (GMRT). \Fig~\ref{fig:light_curves} shows the accumulated light curves from all observations, with maximum flux densities ranging from 0.1 to 10\,Jy. The pulse widths range between 30 and 300\,s, and can appear anywhere within a $\sim$400-s wide window. The morphologies and brightnesses of the observed pulses change dramatically with time, sometimes disappearing below detectability, even for the very sensitive MeerKAT observations. Depending on telescope sensitivity, this nulling fraction is 50--70\,\%. No persistent radio source was detected down to a 3$\sigma$ limit of 60\,$\mu$Jy\,beam$^{-1}$ using MeerKAT UHF data integrated outside of the pulse activity windows.

We measured the arrival times for all the detected pulses spanning nearly 34~years, and using standard pulse timing methods (see Methods), we derive a period $P$ of \srcPlong{}$\pm$\srcPerr{}\,s. Despite the wide variation of pulse arrival times within the pulse phase, this large time lever arm also enables an estimation of the spin-down rate $\dot{P}$, constraining it to be $\lesssim$\srcPdot{}\,\Pdotunit{} (1-$\sigma$ limit). 

\begin{figure}[H]
\centering
\includegraphics[width=89mm]{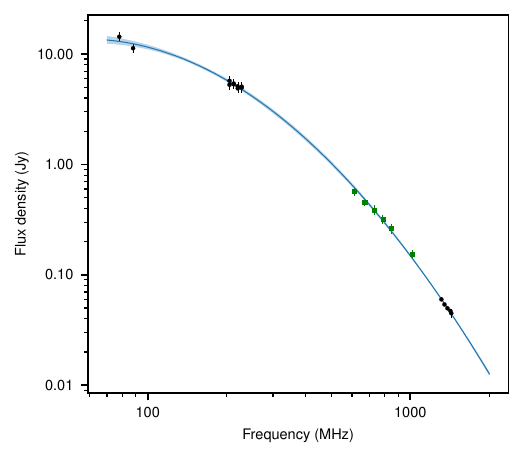}
\caption{Broadband radio flux density spectrum of a typical bright pulse. The data are constructed using pairs of contemporaneous or same-frequency observations taken by the MWA, ASKAP, and Parkes (black points). The blue line shows a curved spectrum (\Eqn~\ref{eq:snu} in Methods) fitted to these points.
The brightest pulse observed by MeerKAT is shown with green squares and has been scaled by a single factor to align it with the rest of the data; these points are not used in the fit.\label{fig:SED}}
\end{figure}

\Fig~\ref{fig:SED} shows the radio flux density spectrum of a typical bright pulse; the radio luminosity of a pulsar-like source generating this pulse is \srcRLum{}\,\ergpers{} (see Methods). This luminosity is larger than the available spin-down luminosity in the neutron star case, similar\cite{2022Natur.601..526H} to \GLX{}. However, this discrepancy has previously been seen in magnetar single radio pulses\cite{2022ApJ...940...72R}. 

\begin{figure}[H]
\centering
\includegraphics[width=179mm]{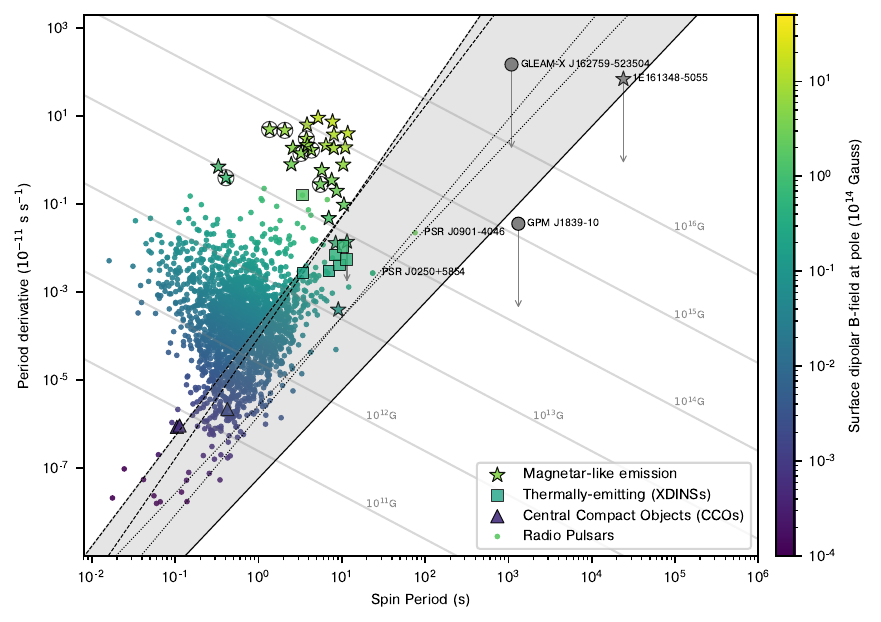}
\caption{Spin period $P$ against period derivative $\dot{P}$ for neutron stars and long-period radio transients.
$\dot{P}$ for \src{} is the 1-$\sigma$ upper limit (see \EFig~\ref{fig:ppdot_search}).
We report all ATNF isolated radio pulsars\cite{2005AJ....129.1993M}, pulsars that show magnetar-like X-ray emission (stars, circled in grey if radio-loud) -- including the radio-quiet long-period magnetar 1E~161348-5055\cite{2006Sci...313..814D,2016ApJ...828L..13R,2016MNRAS.463.2394D}), X-ray Dim Isolated Neutron Stars (XDINSs, squares) and Central Compact Objects (triangles)\cite{2014ApJS..212....6O, 2018MNRAS.474..961C}.
The long-period radio transients are shown with grey circles. The color bar reflects the surface dipolar magnetic field at the pole assuming spin-down due to dipolar losses (see Methods).
The dashed lines correspond to the theoretical death-lines for a pure dipole\cite{1993ApJ...402..264C,2000ApJ...531L.135Z}, dotted lines for a twisted dipole\cite{1993ApJ...402..264C,2000ApJ...531L.135Z}, and the solid lines for the twisted multipole configuration\cite{1993ApJ...402..264C}. The shaded region is the death valley for neutron stars considering all these different lines (see text and Methods for details).}
\label{fig:p_pdot}
\end{figure}

Coherent radio emission from rotating neutron stars has been explained by efficient pair production in the magnetosphere triggered by an accelerating region in the polar cap. Previous emission models have considered curvature radiation or inverse Compton scattering photons as the source of pairs; dipolar, multipolar and twisted magnetic fields\cite{1975ApJ...196...51R,1993ApJ...402..264C}; and vacuum gap or space-charge-limited flows\cite{2000ApJ...531L.135Z}. Considering all these effects, there is a range of parameters (known as ``death valley'') below which pair production is no longer an efficient mechanism and coherent radio emission is no longer expected.
In \Fig~\ref{fig:p_pdot} we plot the period derivative as a function of the spin period for different classes of isolated neutron stars\cite{2005AJ....129.1993M,2014ApJS..212....6O, 2018MNRAS.474..961C}. To encompass the largest parameter range, we have overplotted several ``death valleys'' considering a few extreme cases (see Methods).
\src{} falls at the very edge of the most generous death-line.
Based on our constraints on $\dot{P}$, a classical coherent radio pulsar emission from a rotating neutron star is barely viable.

The long-lived activity of \src{} is extremely puzzling. The radio activity of the $P=$18.18-minute transient \GLX{} was short-lived, persisting for only three months (with no detected nulling episodes) in eight years of observations, with no further detections in archival data or recent monitoring\cite{2022Natur.601..526H}. This led to the postulation that the radio emission could have been powered by a temporary rearrangement of its magnetic fields, preceded by an (unobserved) high-energy outburst, similar to canonical magnetar radio emission\cite{2010ApJ...721L..33L}. \PSR{}, a postulated ultra-long period neutron star pulsar with a rotational period of $\sim$76\,s, has been continuously active since its discovery, but its well-constrained $\dot{P}$ places it above the death lines for several pulsar emission mechanism models\cite{2022NatAs...6..828C} (see also Fig~\ref{fig:p_pdot}). In contrast, \src{} has been active for at least three decades, although with a 50-70\% nulling fraction. This new source has neither a short enough period to be explicable by canonical radio pulsar emission nor a short enough activity window for its radio emission to resemble a typical magnetar outburst.

An observation with \emph{XMM-Newton} simultaneous to the ASKAP observation did not detect X-ray emission from the position of \src{}. \EFig~\ref{fig:xray_limits} shows the limits on the persistent luminosity in the soft X-rays assuming both thermal and non-thermal spectral shapes. For a reasonable range of parameter values, the derived values lie in the range $(0.1-1.5) \times10^{32}$\,\ergpers{}. Two very bright radio pulses were recorded by ASKAP during the X-ray observation, with no X-ray counterparts (see \EFig~\ref{fig:radioxray_bursts}). We derived limits on the X-ray luminosity during the bright radio pulses of the order of $<2\times10^{33}$\,\ergpers{} (see Methods). These limits are orders of magnitudes below those measured for magnetar-like X-ray bursts. These results rule out a direct connection between these radio bursts and magnetar-like X-ray events, as in the case of SGR\,1935$+$2154\cite{2020Natur.587...54C}.
\src{} was also observed in the infrared $K_s$-band with the Espectr\'ografo Multiobjeto Infra-Rojo (EMIR) mounted on the 10\,m Gran Telescopio Canarias (GTC). A faint source was marginally detected at $K_s=19.7\pm0.2$ (see \EFig~\ref{fig:NIR_image}).
However, the present data does not allow us to firmly conclude whether the source is single or blended, or if it is truly associated with the radio source.

There are several alternatives to the neutron star (radio pulsar or magnetar) interpretation. A rotating and highly-magnetic isolated white dwarf, with its larger moment of inertia\cite{2022Ap&SS.367..108K}, could produce pulsar-like radio emission. However, it is perhaps surprising that no close-by highly-magnetic white dwarfs have been observed to produce such emission. The only known white dwarf pulsar emitting in radio, AR Sco, has a $P=$2\,min spin period and a binary companion in a 3.5-hr orbit \cite{2016Natur.537..374M}. However, this radio emission, partly produced by the interaction with the companion star, is three orders of magnitude less luminous than the emission from \src{}. A highly-magnetic rotating proto-white dwarf (or ``sub-dwarf'') is a possibility, and it could potentially be obscured by its red giant progenitor\cite{2022RNAAS...6...27L}; both obscured and unobscured cases could be tested with further optical and IR observations. Low-frequency radio emission has also been detected from star -- exoplanet interactions\cite{2020NatAs...4..577V,2021NatAs...5.1233C} and brown dwarf binaries\cite{2023arXiv230101003V}, with modulations on both rotational and orbital periods. However, such emission is significantly circularly polarised, and is around eight orders of magnitude less luminous than the emission observed from \src{}.

The detection of \src{} and \GLX{} confirms that ultra-long period radio sources are not staggeringly rare, and there will be many opportunities in upcoming Galactic plane surveys to find further examples. Correlation of the observed population with predicted spatial distributions may yield further insights. Higher-frequency radio observations will be less affected by scattering, but if the radio flux density spectrum of these sources is identical, then detecting them beyond 1.4\,GHz will be extremely challenging. Classical coherent radio emission mechanisms involving an accelerating potential gap and subsequent pair-production do not seem viable to explain \src{} (see Rea et al. in prep for further discussion, including the magnetic white dwarf scenario). 
Intriguingly, the persistence of \src{} over three decades indicates that there may be many more long-lived sources, some of which may await discovery in existing archives. Sources similarly active for decades, but with smaller pulse windows, would yield better constraints on $\dot{P}$, and thus even stronger tests of models of coherent radio emission from astrophysical sources.

\begin{methods}

\subsection{MWA}

The Murchison Widefield Array is a low-frequency radio telescope operating in the Murchison region of Western Australia\cite{2013PASA...30....7T, 2018PASA...35...33W}. A Galactic Plane Monitoring programme (project code G0080) was executed during the 2022-A semester, with observations at 185--215\,MHz covering $|b|<15^\circ$ and $284^\circ<l<73^\circ$, for 30\,min integration on a bi-weekly cadence (Hurley-Walker et al. in preparation will describe the survey in full). Following the detection of \src{}, Director's Discretionary Time observations (code D0038) were used to monitor the evolution of the source every $\sim$48\,hr from 2022-07 until 2022-09. The observations were 5\,min long and in five 30.72-MHz bands iterating through 72--231\,MHz, yielding a resolution of $5'$--$45''$ and snapshot noise levels of $150$--$25$\,mJy\,beam$^{-1}$. Initial images were formed from the relevant G0080 pointings and all D0038 observations using the GaLactic and Extragalactic All-sky MWA -- eXtended (GLEAM-X) pipeline\cite{2022PASA...39...35H}, and placing a mask at the location of \src{} to avoid including it in the deconvolution model. Subsequently, the data were re-imaged with the continuum sources subtracted, on a 4-s time cadence. If \src{} was detected, the model-subtracted imaging was repeated forming small images around \src{} at an 0.5-s time cadence and with 384$\times$80\,kHz, 192$\times$160\,kHz, 96$\times$320\,kHz, or 48$\times$640\,kHz~channels for central frequencies at 88, 118, 154, or $>$180\,MHz respectively, in order to avoid dispersion smearing within the channels. Dynamic spectra were formed by measuring the (primary-beam-corrected\cite{2015RaSc...50...52S}) flux density at the location of \src{} in each time/frequency voxel.

\subsection{Parkes}

On UTC 2022-07-17 (2459778 BJD-TDB), we ran commensal observations of \src{} with the MWA and the 0.7--4.0 GHz receiver\cite{Hobbs:2020} of the CSIRO Parkes `Murriyang' telescope. Dynamic spectra were recorded in filterbank format (32-bit float) using the Breakthrough Listen data recorder\cite{Price:2018,Lebofsky:2019,Price:2021}, with 0.5\,MHz spectral resolution and 100\,$\mu$s time resolution. Due to strong radio interference at lower frequencies, and decreasing pulse S/N at higher frequencies, we selected the band 1.090--2.365\,GHz for further analysis. 
Before observing \src{}, we observed calibrator source PKS\,1934$-$638 and applied standard position-switching techniques\cite{Winkel:2012} to determine the system temperature, using a log-polynomial flux model for  PKS\,1934-638\cite{Reynolds:1994}. We used off-pulse spectra when pointing toward \src{} to account for the additional system temperature contribution from the Galactic plane, then fit a 7-order polynomial to describe the system temperature across 1.090--2.365\,GHz. To calibrate pulses from \src{}, we divided the uncalibrated on-pulse spectra by the mean off-pulse spectra, then applied the system temperature correction detailed above. Before analysis, bands with known radio interference were then flagged, and data were downsampled to 0.1\,s. 

Data from \src{} were also recorded in 8-bit PSRFITS format using the standard Parkes signal processor, Medusa. However, as Medusa applies automatic gain correction (AGC) on similar timescales to the \src{} pulse length, detection of pulses in these data is difficult. We note that AGC and red noise introduced by receiver gain fluctuations are confounding factors that may explain why objects similar to \src{} have not been found previously in data from single-dish pulsar surveys.

\subsection{MeerKAT}
MeerKAT is a large radio interferometer operated from the South African Radio Astronomy Observatory\cite{2016mks..confE...1J}. Observations were made under project code DDT-20220718-NH-01 comprising four hours of time on source at UHF (500--1060\,MHz); the correlator ran with 1-MHz channels and 2-s integration.
The data were also recorded by the Accelerated Pulsar Search User Supplied Equipment (APSUSE) and the Pulsar Processor (PTUSE) simultaneously. The intensity data of seven beams were recorded on the APSUSE as filterbank format with a time resolution of $\sim$30\,$\mu$s and 1024~frequency channels. The PTUSE recorded the full polarization data of the central beam as PSRFITS search-mode data, which has a time resolution of $\sim15$\,$\mu$s and 1024~frequency channels. Unfortunately, due to low position accuracy at the time of observing, none of the coherent beams were directly on-source, causing lower signal-to-noise, baseline variation, and artificial frequency steepening in the high time resolution data (the correlator data were unaffected).

The observatory calibration solutions were used to calibrate the correlator data, RFI flagging was performed with AOFlagger\cite{2012A&A...539A..95O}, and imaging was performed with \textsc{WSClean}\cite{2014MNRAS.444..606O}. Similarly to the MWA imaging, a mask was placed at the location of \src{} and a deep image was formed of all the data. The position of the source was measured to be RA$=$18$^\mathrm{h}$39$^\mathrm{m}$02.00$^\mathrm{s}$, Dec$=-10^\circ$31$'$49\farcs37, with a 1-$\sigma$ uncertainty of 0\farcs15.  Within the pulse window of the source, the data were re-imaged with the deep model subtracted, on a 2-s cadence, across 76~frequency channels spanning 568--1053\,MHz. Dynamic spectra were formed by measuring the flux density of \src{} in each time/frequency voxel. Data outside of the pulse window was re-imaged, cleaning down to $3\times$ the local RMS noise $\sigma$, and down to $1\sigma$ on any pixel previously found to contain flux density (auto-threshold 3, auto-mask 1). No persistent radio source was detected down to a 3-$\sigma$ limit of 60\,$\mu$Jy (at a central frequency of 812.5\,MHz). Polarisation mesaurements were made of the first pulse using the brightest pixel at the location of the source in each Stokes parameter.

For the high time resolution data, we firstly searched for single pulses on the PTUSE data with \textsc{TransientX}\cite{TransientX}. We used two sets of search parameters: (1) To search for narrow pulses, we used a maximum search width of 100\,ms and a DM range from 0 to 500\,\DMunit{}; (2) To search for wide pulses, we used a maximum search width of 10\,s and the same DM range with the data downsampled to 10\,ms. As a result, a dozen of pulses with pulse width larger than 100\,ms and a $\sim$30\,ms pulse were detected within the two 120-s long pulses at UTC 2022-07-20T19:12:33 and 2022-07-20T19:35:43. We then chopped the data with the time spans that contained pulses in the two periods and formed PSRFITS fold-mode data for further investigation. The data recorded on the PTUSE was already calibrated\cite{2021MNRAS.505.4483S}; we performed RM synthesis on the chopped data with \textsc{RMfit} in \textsc{PSRCHIVE}\cite{2012AR&T....9..237V}, which gave the RM estimation of $-531.83\pm 0.14$ and $-532.2\pm2.2\ \mathrm{rad\ m}^2$ in two bursts, respectively, as shown in \EFig~\ref{fig:rmfit}
. It should be noted that the pulsed emission lasted for $\sim120$\,s in the first period, so the data is strongly affected by the baseline variation. However, the baseline variation does not affect RM synthesis, because it can be viewed as a DC component in the RM spectrum. Due to this effect, we only measured the total intensity I, linear polarized intensity L, circular polarized intensity V and position angle (PA) -- e.g. \EFig~\ref{fig:polarization} shows a $\sim$25-ms duration pulse.

To increase S/N, we downsampled the APSUSE data to $\sim3.9$\,ms, and removed the baseline in the data of the central beam based on the baseline variation of the data in other beams, which yields the dynamic spectrum of the $\sim120$\,s pulse shown in
\EFig~\ref{fig:dynspec}. The ``DM Phase'' structure maximisation algorithm\cite{2019ascl.soft10004S} is applied to these data, estimating a DM of \srcDM{}$\pm$2.5\,\DMunit{}. All subsequent analysis uses this DM for de-dispersion.

\subsection{ASKAP}

The source was observed with the ASKAP radio telescope\cite{2021PASA...38....9H} as part of a target-of-opportunity observation within the Variables and Slow Transients survey (VAST) \cite{Murphy2013}. ASKAP was pointed at (RA$=$18$^\mathrm{h}$39$^\mathrm{m}$02$^\mathrm{s}$0, Dec$=-10^\circ$31$'$49\farcs5) using a $closepack36$ footprint \cite{2021PASA...38....9H} with 0.9 degree pitch (spacing between beams) with a central frequency of 1367.5 MHz and a bandwidth of 144 MHz (288 MHz was observed but the lower half of the band is discarded as a result of GNSS-based RFI). This places beams 15 and 21 closest to the target source. Observations were performed under the alias AS113$\_$65 and fell within scheduling blocks 43636 (starting 2022-09-03 07:35:02, 2 hours total) and 43703 (starting 2022-09-04 12:02:43, 5 hours total) for the first observing epoch and 44023 (starting 2022-09-14 07:33:25, 6~hours total) for the second epoch. The first two observations were concatenated (7~hours total) to improve convergence of deconvolution (as complex structures from the Galactic Plane caused problems with the limited sampling of the $(u,v)$-plane).

Each observation was calibrated and imaged using the default ASKAPsoft pipeline \cite{2017ASPC..512..431W}. The flux density scale, bandpass, on-axis leakage, and initial gain calibration are derived from observations of PKS\,1934$-$638. The ASKAPsoft pipeline produces calibrated visibilities, Stokes~I and Stokes~V images, and Stokes I source catalogues for each observation. All data products are made available through the CSIRO ASKAP Science Data Archive \cite{casda,Huynh2020} under scheduling blocks 43696 (first epoch with concatenated observations) and 44023 (second epoch).

For polarisation investigations, the deconvolution model was subtracted from the visibility data using the ASKAPsoft tool \emph{contsub}. The visibility data was then phase-shifted to the location of the source and averaged across all baselines to produce dynamic spectra for each of the instrumental polarisations. These were then converted to Stokes I, Q and U. Rotation measure synthesis and cleaning was then performed using the \emph{RMsynth1d} and \emph{RMclean1d} scripts from RM-tools\cite{2020ascl.soft05003P}. At the Stokes I peak (2022-09-14 11:08:18.485) we find a peak polarised intensity of $13.0\pm1.2$\,mJy\,beam$^{-1}$\,RMSF$^{-1}$ at a Faraday Depth of $-532.7\pm17.3$\,rad\,m$^{-2}$ with an associated Stokes I peak of $361.8\pm2.2$\,mJy\,beam$^{-1}$ (a fractional polarisation of 3.5\,\%). For integration immediately following the Stokes I peak (2022-09-14 11:08:28.438) we find a peak polarised intensity of $61.8\pm1.3$\,mJy\,beam$^{-1}$\,RMSF$^{-1}$ at a Faraday Depth of $-531.3\pm4.0$\,rad\,m$^{-2}$ with an associated Stokes I peak of $102.7\pm2.2$\,mJy\,beam$^{-1}$ (a fractional polarisation of $\sim$60\%).

\subsection{ATCA}
The Australia Telescope Compact Array (ATCA) performed observations of the position of \src{} on 2022-07-17 under proposal identification C3451. The observations were conducted using the 4\,cm dual receiver with frequencies centred at 5.5 and 9\,GHz, each with a 2\,GHz bandwidth. The array was in the hybrid configuration H214. Removing antenna CA06, which provided $\sim4$\,km baselines, the longest baseline with the remaining 5 antennas was 247\,m, resulting in a resolution of $\sim45$\,arcsec. These data were processed using the radio reduction software Common Astronomy Software Applications (CASA) package \cite{casa_2007}, version 5.1.2 using standard techniques. Flux and bandpass calibration was performed using PKS\,1934--638, with phase calibration using interleaved observations of 1829-106.

The observations were scheduled to be simultaneous with MWA and Parkes observations between 2022/07/17 09:42:58.4 and 2022/07/17 15:42:19.6 UT, totalling $\sim$4\,h on source. 
No persistent radio source was detected at the position of \src{} at 5.5\,GHz with a $3\sigma$ upper limit of 0.1\,mJy\,beam$^{-1}$. 
Assuming a DM of \srcDM{}\,\DMunit{}, then the pulse detected by MWA at 2022/07/17 12:53 at 185\,MHz (Fig.~1) would have arrived 33\,s earlier at 5.5\,GHz.
Based on the observed pulse width seen by the MWA, we extracted a 150-s integration between 2022/07/17 12:52:30 and 12:55:00, yielding a $3\sigma$ upper limit of 0.6\,mJy at 5.5\,GHz. Scaling this limit by the ratio between the flux density of the MWA observed 185\,MHz burst and the expected flux density in the formed spectrum at 185\,MHz (a factor of $\sim$10), we arrive at a $3\sigma$ upper limit of 6\,mJy at 5.5\,GHz.

\subsection{VLITE}
The VLA Low-band Ionosphere and Transient Experiment (VLITE) \cite{2016ApJ...832...60P,2016SPIE.9906E..5BC} is a commensal instrument on the NRAO VLA that records and correlates data across a 64 MHz bandwidth at a central frequency of $\sim$340 MHz. VLITE operates on up to 18 antennas during nearly all regular VLA operations since 2017-07-20, accumulating roughly 6000 hours of data per year. All VLITE data are processed through an automated VLITE calibration and imaging pipeline\cite{2016ApJ...832...60P}, producing final calibrated visibility data sets and self-calibrated images. These VLITE images and associated META data are then passed through the VLITE Database Pipeline (VDP)\cite{2019ASPC..523..441P} to populate a Structured Query Language (SQL) database containing cataloged sources. 

We searched VDP for archival images where \src{} lies within 2$^\circ$ of the pointing center of VLITE observations taken in the VLA A and B configurations. We note that the half-power radius of the VLITE primary beam response is $\sim$ 1.25$^\circ$ but the system is sensitive to sources well beyond this. We identified 86 VLITE data sets recorded between 2017-11-02 and 2022-06-29 that matched the search criteria. For this paper, we concentrated primarily on A configuration observations which provide the highest angular resolution ($\sim$5$^{\prime\prime}$) and sensitivity (few mJy\,beam$^{-1}$ at the field center), re-imaging 42~data sets, but also included a further 11~observations from B-configuration (resolution $\sim$20$^{\prime\prime}$) in an effort to expand the detected burst time range. 

Among the 53~data sets identified above, there are no targeted observations of \src{}, rather the target position ranges from 1.1$^\circ$ to 2.0$^\circ$ degrees away from the phase center of the VLITE observations. For each VLITE data set, the self-calibrated visibilities were first phase-shifted to the position of \src{} using the tool \emph {chgcentre} and then \textsc{WSClean}\cite{2014MNRAS.444..606O} was used to produce a time series 
of images of the target down to 2-s intervals (i.e., the data integration time) for A configuration and 6-s intervals for B configuration. The primary-beam corrected noise in the 2s snapshot images at the position of \src{} (i.e.\ far out from the VLITE phase center) is on average 204\,mJy\,beam$^{-1}$ (A-configuration) while the primary-beam corrected noise (again far out from the VLITE phase center) in the 6-s snapshots is on average 160\,mJy\,beam$^{-1}$ (B-configuration). \src{} was detected by VLITE in four datasets from 2022 (one in 2022-04 and three in 2022-05), one dataset from 2021-02, and one pulse was split across two observations on 2018-04-21 (\Fig~1).

\subsection{GMRT}

The archives of Giant Metrewave Radio Telescope were searched for data in which \src{} would lie within the field-of-view. PSR~J1835$-$1020, 47\farcm3 away, was observed in P-band on 2015-01-03 and 17, each in three 20--25-min scans with 2-s integrations. The spectral bandwidth was split into 256~channels each 130.2083\,kHz wide, covering 33.333\,MHz with a central frequency of 322.6\,MHz. Detections were also found in observations using 16.8-s integrations on 2007-11-17 and 29, and 2007-12-14 toward G21.5$-$0.9, 1\fdg4 away, and on 2002-05-12 and 13 toward GS\,022$-$02, 9\farcm9 away, in $\sim$1\,hr scans in 2007, and in $8\times10$-min scans about 1\,hr apart in 2002. The spectral bandwidth of each observation comprised 64 125-kHz channels spanning 8\,MHz, at a central frequency of 240\,MHz.

Calibration and imaging was performed in the Astronomical Imaging Processing System (AIPS\cite{AIPS}). The amplitude and phase calibrations were performed using 3C286, J\,1830-360, and J\,1822-096. The field was self-calibrated on a widefield image of 19~facets covering an aperture $1^\circ$ in radius. Both a deep image comprising all data and single-integration snapshots with an inner $(u,v)$ cut of 2\,k$\lambda$ (2017), 1\,k$\lambda$ (2007), and 0.5\,k$\lambda$ (2002) were created. The flux density scale was adjusted to match the deep-field measurements of the flux densities of sources in the field to catalogue values\cite{2018MNRAS.474.5008D}, and correction factors 2.25, 1.0, and 1.61 were applied to the 2015, 2007, and 2002 snapshot images. The RMS noise of the 2-s snapshot images was typically 25--50\,mJy\,beam$^{-1}$ before corrections for primary beam attenuation, and the resolution was $20\farcs1 \times 7\farcs6$. Correction factors of 2.29 (2015), 4.10 (2007), and 1.02 (2002) were applied for the primary beam attenuation at the source location.


\subsection{VLA}

The archives of the Very Large Array were searched in a similar fashion, finding two $\sim$325\,MHz, 10-s resolution observations of target ``G21'', 1\fdg3 away. One observation was taken on 2001-04-28 in B-config over 07--15\,h in 9$\times$30-min scans at 321.6 and 327.5\,MHz (with 31 channels spanning 3.03\,MHz bandwidth for each IF), while another was taken on 2001-08-31 in C-config via 3$\times$1-hr scans from 02--05\,h at 327.5\,MHz with 3.03\,MHz bandwidth. A third 10-s resolution observation of target "20-20", 2\fdg0 away, was made in D-config on 1988-09-23 and 24, consisting of three 20-min, single channel, 3.1 MHz wide scans at 327 and 333 MHz. The data were processed in AIPS, using the amplitude and phase calibrators 3C286, 3C48, 1822-096, and 1828+487. Similar processing steps were performed as with the GMRT, performing faceted self-calibration on a widefield image and subsequently imaging the snapshot images with an inner $(u,v)$ cut of 0.1--1\,k$\lambda$. The RMS of the 10-s snapshot images before corrections for primary beam attenuation was typically 50\,mJy\,beam$^{-1}$ (B-config), 100\,mJy\,beam$^{-1}$ (C-config), and 125\,mJy\,beam$^{-1}$ (D-config), and the resolution of the images was $\sim$30$''\times15''$ (B-config), $\sim$90$''\times60''$ (C-config), and $\sim$400$''\times200''$. Correction factors of 2.0 (2001) and 6.0 (1988) were applied for the primary beam attenuation at the source location.

\subsection{\textit{XMM-Newton}}

The \emph{XMM-Newton} satellite observed the field of \src{} on 2022-09-14, starting at 04:18:59 UTC and ending at 12:20:55 UTC (Id. 0913991101; satellite revolution \#4169). The observation lasted $\simeq$28.9\,ks, completely overlapping the ASKAP observation. 

The EPIC-pn\cite{struder01} was configured in the Large Window mode, whereas both MOS instruments\cite{turner01} operated in Full frame mode. The data were processed and analysed using the Science Analysis System\cite{gabriel04} (\textsc{sas}; v. 20.0). Visual inspection of the time series for the entire field of view in selected energy bands revealed the presence of strong background flares. These were removed from the data by applying ad hoc intensity thresholds on the time series, resulting in effective on-source exposures of $\simeq$17.6\,ks for the EPIC-pn and $\simeq$22.6\,ks for both EPIC-MOS cameras.

No X-ray emission is detected at the position of the radio source. We adopted an aperture radius of 20\arc\ for the source photons and an annular region centred on the source with inner and outer radii of 20\arc\ and 40\arc\ to estimate the background level, and computed a 3-$\sigma$ upper limit on the net count rate of $0.002$\,counts\,s$^{-1}$ in the 0.3--10\,keV energy band. 
This upper limit was then converted into luminosity limits by assuming an absorbed blackbody model or an absorbed power-law model for the X-ray spectral shape. We adopted the absorption column density estimated in the direction of the source using the HI4pi survey over a 0.1 deg$^2$ around the source position: $\nh\sim5.5\times10^{21}$\,cm$^{-2}$. We evaluated the limits on the source X-ray luminosity by varying the blackbody temperatures in the range $kT=0.1-0.9$\,keV in steps of width $\Delta kT = 0.02$\,keV for the first model, and the power-law photon index in the range $\Gamma=1-4$ in steps of width $\Delta \Gamma = 0.1$ for the second model, and assuming a distance of 5.7\,kpc. \EFig~\ref{fig:xray_limits}
shows the interpolation curves that best approximate the limits thus obtained. The allowed parameter space on these planes, taking into account uncertainties in the value of the source distance, is indicated using shaded regions.

We also performed a dedicated search for X-ray emission during the two time windows in which the radio brightness at the source position measured in the simultaneous ASKAP observations was at least three times larger than the RMS noise of the image. Again, no X-ray emission was detected in either time window. The inferred 3-$\sigma$ upper limits on net count rate were 0.07 and 0.08\,counts\,s$^{-1}$ (0.3--10\,keV) during the first and the second radio flaring event, respectively. Assuming an absorbed power-law spectrum with the above-mentioned \nh\ and $\Gamma = 2$, these values convert into X-ray luminosities of $L_{\rm X, Flare1} < 1.8\times10^{33}$\,\ergpers{} and $L_{\rm X, Flare2} < 2.0\times10^{33}$\,\ergpers{}.

We note that an X-ray source 
is detected near the position of \src{}, with a net EPIC-pn count rate of 0.028$\pm$0.002\,counts\,s$^{-1}$ over the 0.3--10\,keV energy range. Its position, calculated with a maximum-likelihood source detection pipeline, is 
R.A. = 18$^\mathrm{h}$38$^\mathrm{m}$57$\farcs$343, Decl. = $-10^{\circ}$29$^{\prime}$25$\farcs$6 (J2000.0; the uncertainty on the position is 0.4\arc\ at a confidence level of 1$\sigma$). This is $\simeq$2.6 arcmin away from the nominal position of \src, ruling out a possible association between this X-ray source and the radio source.
Following the naming convention for newly discovered X-ray sources, we denote it as XMMU\,J183857.4$-$102926. For the sake of completeness, we have performed a spectral analysis of the X-ray emission from this source. The background-subtracted EPIC-pn spectrum is well described by an absorbed power-law model with $\nh=(1.7\pm0.5)\times10^{22}$\,cm$^{-2}$ and $\Gamma=1.2\pm0.2$ (reduced chi squared of $\chi^2_r=0.48$ for 11 degrees of freedom). Its X-ray flux is $3.5_{-0.3}^{+0.4}\times10^{-13}$\,\flux{} (0.3--10\,keV), after correcting for absorption effects. The inferred \nh\ value is larger than the expected Galactic value along the direction of the source ($\nh_{,\rm Gal}=7.5\times10^{21}$\,cm$^{-2}$)\cite{willingale13}, pointing to either an intrinsically absorbed Galactic source or an extra-galactic source.

Six images with a duration of 4--4.5\,ks were collected fully overlapping with the X-ray data using the Optical/UV Monitor telescope\cite{mason01} on board \emph{XMM-Newton}, each using a different filter. The $UVW2$ filter (effective wavelength of 212\,nm) was in place during the epochs when the two flares were detected in the radio observation. No emission was detected from the position of \src\ in any image, down to the following 5-$\sigma$ limiting magnitudes: $V>19.8$\,mag, $U>20.6$\,mag, $B>21.0$\,mag, $UVW1>20.2$\,mag, $UVM2>19.2$\,mag
and $UVW2>18.3$\,mag.

\subsection{Grantecan}
The Espectr\'ografo Multiobjeto Infra-Rojo (EMIR\cite{garzon22}) mounted on the Naysmith-A focus of the 10.4-m Gran Telescopio Canarias observed the field of \src\ twice in October 2022. The first observation began on October 25 at 19:48:52 UTC and lasted $\simeq$1.5\,hr, while the second one began on October 26 at 19:26:56 UTC and lasted $\simeq$1.4\,hr. Both observations were carried out in service mode under clear sky and dark Moon conditions.
Each night, 480 frames each with an exposure of 3\,s were acquired in the $K{\rm s}$ filter (central wavelength of 2.16\,$\mu$m and full width at half maximum of 0.312\,$\mu$m), using the standard seven-point dither pattern. The EMIR images were processed with custom software to produce the final stacked image, shown in \EFig~\ref{fig:NIR_image}.

The stacked image was reduced and analyzed using IRAF's DAOPHOT/ALLSTAR tool. First, we checked the astrometric solution of the working image by comparing the centroids of the fitted stars with their positions in the Gaia DR3 catalog. We then transformed the ALLSTAR catalog (in XY coordinates) into a WCS catalog using the IRAF task WCSCTRAN. The average difference between the EMIR pipeline astrometric solution and the Gaia DR3-based solution was -0.03 arcseconds (-0.01 pixels) and -0.02 arcseconds (-0.01 pixels) in the region near \src.

We then ran DAOPHOT/ALLSTAR using a quadratic PSF model of bright stars (FWHM = 4 pixels) to perform forced photometry at the image position of \src, obtained from the improved Gaia DR3 astrometric solution. We repeated this process several times to check the robustness of the fit. The differences between the optical fitted positions and the radio positions were found to be within 0.61 pixels and 0.52 pixels in the X and Y image coordinates, respectively, corresponding to 0\farcs11 and 0\farcs10 in RA and DEC. The optical source is therefore found at: RA(J2000)= 18h39m01.992s and DEC(J2000)=-10d31m49.60s at a significance level of 4.1$\sigma$.

We also performed an additional check by subtracting the image obtained with the PSF-fitted stars from the original image, resulting in negative values at the position of \src. The instrumental magnitude at that position is 19.23$\pm$0.20. However, the sharpness parameter, which measures how well the PSF fits the object, was highly negative (sharpness = -3), indicating that the source is either extended or blended.

The $K_s$-band photometry, calibrated using 2MASS stars in the EMIR field of view, gives for the detected source a magnitude of $K_s$=19.73$\pm$0.2 (measured) $\pm$0.12 (calibration). Based on a distance of 5.7$\pm$2.9\,kpc derived from DM measurements, and correcting for interstellar reddening in the $K_s$ band ($A_K$=0.683; \cite{schlafly11}), the absolute magnitude of the source is calculated to be $K_{\rm abs}= 5.3_{-0.9}^{+1.5}$. If the optical source is a single star on the main sequence, it would be of spectral type ranging from mid $K$ -- mid $M$ \cite{bilir08}.

\subsection{Pulse polarisation and substructure}

Polarisation was detectable with both ASKAP and MeerKAT. The rotation measure (RM) was derived from the two highest S/N $\sim$30-ms bursts observed by MeerKAT as $-531.83\pm 0.14$ and $-532.2\pm2.2$\,\RMunit{}, as shown in \EFig~\ref{fig:rmfit}. The degree of linear polarization of the fine-timescale structure is $\sim$100\% (\EFig~\ref{fig:polarization}). Analysis of the correlator data over the same full $\sim$120-s pulse (\Fig~\ref{fig:meerkat_pol}) yields consistent results, with peak RMs in each timestep typically $-533\pm2$\,\RMunit{}. A weak ($\lesssim$10\,\%) circular polarisation component is also detected in several timesteps of this pulse, and several others observed with ASKAP and MeerKAT.

In the high S/N pulse shown in \Fig~\ref{fig:meerkat_pol}, the polarisation position angle is generally flat. The high-time resolution data reveals sudden jumps of 90$^\circ$ in polarisation phase angle, lasting from 0.2--4\,s.
The secondary mode is also typically coincident with weaker Stokes~I. These mode switches resemble the orthogonal polarization modes (OPMs) common in pulsars\cite{Ekers1969}. Although the origin of OPMs in pulsars is still debated\cite{Johnston2017}, their presence in radio pulses from \src{} could imply a neutron-star-like emission mechanism.

The MeerKAT data at 3.9\,ms (\EFig~\ref{fig:dynspec}) reveals fine pulse substructure, with features as narrow as the time resolution. Quasi-periodic oscillatory (QPO) features are seen in many of the pulses. We performed an autocorrelation lag analysis on well-resolved pulses, finding a range of characteristic timescales, with almost every pulse exhibiting particularly significant oscillations at $\sim$50\,s (see \EFig~\ref{fig:acf}). We extended this to a cross-correlation analysis between pulses separated by one or two periods, and found some morphological correlation; lags of zero and $\sim$50\,s were particularly preferred, but this was not always consistently observed.
In pulsars, the first non-zero lag peak indicates the characteristic separation between sub-pulses\cite{1979AuJPh..32....9C}.
The features we see are on scales of $\sim$15--75 milliperiods, one to two orders of magnitude larger than observed in pulsar microstructure\cite{2015ApJ...806..236M}, i.e. they do not follow the linear trend between QPO scale and period observed in canonical pulsars. 
However, the observed changes in burst structure and fluence (\Fig~\ref{fig:light_curves}, \EFig~\ref{fig:flux_hist}) point to highly dynamic emission environments, temporal modulations of which could potentially cause the QPO features that are seen.

\subsection{Period and Period Derivative}

The light curves from all telescopes were de-dispersed to \srcDM{}\,\DMunit{} and the timestamps shifted to reflect the arrival time of a zero-DM signal. Since many of the archival detections were made serendipitously in observations that were not set up to specifically observe this source, not all pulses had complete sampling across the on-pulse region.
Because of these incomplete pulses, the wide (30--300\,s) on-pulse region, the diversity of morphology of the pulses, we estimated the time of arrival (TOA) of each pulse by smoothing the light curves with a conservatively wide Gaussian kernel ($\sigma = 300\,$s) and defining the TOA to be the peak time bin of the smoothed signal.

The uncertainties on the TOAs are critical for the timing analysis, and since the pulses have features within a $\pm200$\,s window, the noise budget is dominated by this ``jitter''. To determine its size, we use the 2022 data where the light curves are fully sampled. For each pulse, we measured the offset between the measured (barycentred, de-dispersed) TOA and its expected arrival time in the case of a perfect pulse. We plotted a histogram of these residuals and found a roughly Gaussian distribution, measuring a standard deviation of 38\,s. For fully-sampled light curves, we assign this value as the error of the TOA; for those light curves truncated by observation length or where only one or two peaks are detected due to S/N, we double the error to 76\,s.

The TOAs were then compiled and passed into PINT pulsar timing package\cite{2021ApJ...911...45L} for timing analysis.
PINT corrects for various timing errors associated with each individual telescope systems (e.g., clock errors) and its location at the time of observation.
Among these timing corrections, only the barycentric correction (arising from the location of the Earth in its orbit using DE436 ephemeris for the Solar System barycenter) was found to be significant relative to the TOA uncertainty. 

We performed a brute-force grid search for the spin frequency, $f$, and first spin frequency time derivative, $\dot{f}$, using PINT's gridding utilities. 
First, we fit the TOAs using PINT's Downhill Weighted Least-Squares fitting algorithm where we froze all parameters except $f$, which results in reasonably flat residuals and a reduced $\chi^2$ score of $\chi^2_{\rm best}=1.18$. 
Using this initial fit as a starting point, we searched a spin-frequency range corresponding to $\pm75\sigma_f$, where $\sigma_f$ is the uncertainty in $f$ from the initial fit. This amounted to search the range $f = 0.75861265 \pm 0.00000216\,{\rm mHz}$ with 150~linearly-spaced sample points.
We assumed that the spin frequency derivative was zero initially, but searched between $\dot{f} = 0 \pm 3.78\times 10^{-17}\,{\rm Hz\,s^{-1}}$ with 300~linearly spaced sample points. 
The search range for this parameter was determined by fitting both $f$ and $\dot{f}$ to the data and extracting the resulting uncertainty on $\dot{f}$, which was $\sim 5\times10^{-19}\,{\rm Hz\,s^{-1}}$, and inflating that value by a factor of 75 as in the case of $f$. 
Note that PINT performs a F-test when adding new parameters to the timing model and determined that, with our current data, any improvement in the model from adding $\dot{f}$ is very likely spurious, therefore we can only treat the result as a limit.

For each grid cell that samples the spin frequency and frequency derivative pair ($f_i$, $\dot{f}_i$), we compute the difference in $\chi^2$ scores, $\Delta\chi^2_i = \chi^2_i - \chi^2_{\rm best}$. 
The minima of this $\Delta\chi^2$ grid indicate the nominal best solution(s) for the inclusion of $\dot{f}$ in the timing model. The minimum reduced $\chi^2$ in the grid is $\chi^2_{\rm best}=1.12$ (i.e. adding $\dot{f}$ to the model did not significantly improving the timing).
Confidence intervals are then computed by comparing the $\chi^2$ difference to that expected from a $\chi^2$ distribution with 2 parameters (degrees of freedom).
\EFig~\ref{fig:ppdot_search}
shows the results, with the 3-, 2-, and 1-$\sigma$ confidence intervals indicated with ellipses. The best-fit $\dot{f}$ is consistent with zero, so we are only able to derive lower limits on $\dot{f}$.
To select the most conservative upper limits on the period derivative $\dot{P}$, we use the minimum value of $\dot{f}$ along each of the the 3-, 2-, and 1-$\sigma$ contours.

\subsection{Radio spectrum}
\label{sec:spectral_index}

The brightness and morphology of the pulses of \src{} vary strongly pulse-to-pulse. We thus endeavoured to make several simultaneous measurements of the same pulse at widely separated frequencies. Two key measurements were: a pulse detected at 200--231\,MHz and after delay by the ISM, in the following observation at 72--103\,MHz; and a pulse detected at both the MWA and Parkes, at 200--231\,MHz and 1345--1520\,MHz. Additionally, the bandwidths of the MeerKAT and ASKAP telescopes were sufficient for deriving in-band spectral index measurements across 584--1043\,MHz and 1295--1440\,MHz, respectively. Assuming that the shape of the radio spectrum is constant, and only the brightness varies, we scaled the MWA-Parkes pair to match the MWA-MWA pair via the common measurement at 200--231\,MHz. We also scaled the ASKAP data to match the Parkes data at the common frequency of 1400\,MHz. 

We used the \texttt{scipy} implementation of the Levenberg-Marquardt least-squares fit of a model of the form
\begin{equation}
    S_\nu = S_\mathrm{1\,GHz}\left(\frac{\nu}{\mathrm{1\,GHz}}\right)^\alpha \exp{q \left(\log{\frac{\nu}{\mathrm{1\,GHz}}}\right)^2}
    \label{eq:snu}
\end{equation}
measuring $\alpha=\srcalpha{}\pm\srcalphaerr{}$ and $q=\srcq{}\pm\srcqerr{}$. The flux density at 1\,GHz for this particular dataset is $148\pm3$\,mJy, but the pulses can be brighter, fainter, or not appear at all (see Supplementary Table and \EFig~\ref{fig:flux_hist}). The reduced-$\chi^2$ of the fit is 0.35. The MeerKAT data did not overlap in time or frequency with any other observations so could not be used in the fit, but was scaled by an arbitrary factor (of 13.8). After this scaling, it shows good agreement with the fitted SED, implying that the form of the spectrum does not change dramatically over time.

\subsection{Radio luminosity calculation}

Assuming that the pulses of \src{} are produced via some form of pulsar-like beam intersecting our line-of-sight, we can use geometric arguments to convert the measured flux density to a radio luminosity. For a circular beam with an opening angle that may have some frequency dependence, $\rho(\nu)$, the total luminosity is
\begin{equation}
   L = 4\pi d^2
   \int
   \sin^2{\left[\frac{\rho(\nu)}{2}\right]} S_\nu \,d\nu,
\end{equation}
with distance $d = 5.7\,$kpc. For the fitted log-parabolic spectrum (\Eqn~\ref{eq:snu}), the largest contribution to the flux density is from the frequency range ${\sim}10\,\mathrm{MHz} \lesssim \nu \lesssim 1\,\mathrm{GHz}$. If we assume the beam solid angle over this frequency range can be approximated by a power law with index $\beta$, the luminosity can be expressed
\begin{equation}
     L =  4 \pi d^2 \Omega_\mathrm{1\,GHz} 
     \int \left(\frac{\nu}{\mathrm{1\,GHz}}\right)^\beta S_\nu \,d\nu,
\end{equation}
where $\Omega_\mathrm{1\,GHz}$ is the beam solid angle scaled to $1\,$GHz.

Since contributions from frequencies outside the specified range are negligible for the spectrum \eqref{eq:snu}, we can integrate over all frequencies to obtain a reasonable estimate of the total luminosity,
\begin{equation}
    L \approx 4\pi d^2  S_\mathrm{1\,GHz} \Omega_\mathrm{1\,GHz} \sqrt{\frac{\pi}{q}} \exp \left[ -\frac{(\alpha + \beta + 1)^2}{4q} \right],
\end{equation}
where $\alpha$, $q$, and $S_\mathrm{1\,GHz}$ are the fitted spectrum parameters defined above.

Obtaining an accurate luminosity requires knowledge of the beam geometry parameters ($\Omega$ and $\beta$)\cite{2022MNRAS.514L..41E} which are currently only loosely constrained by the observed pulse duty cycle. However, if we adopt a pulsar-like radius-to-frequency mapping and allow the beam opening angle to be governed by a dipolar magnetic field configuration, then a typical value\cite{2003A&A...397..969K} for the power law index is $\beta = -0.26$. In this case, the luminosity is of order $L \approx \Omega_\mathrm{1\,GHz} \times 10^{28}\,$\ergpers{}. More detailed beam models\cite{1993A&A...272..268G,1993ApJ...405..285R} that can place further constraints on the beam scale factor $\Omega_\mathrm{1\,GHz}$ are not likely to change this estimate by more than an order of magnitude.

\subsection{Error on the distance estimate}

To convert the dispersion measure (DM) into a distance, we use a model of the Galactic electron density, YMW16\cite{2017ApJ...835...29Y}. Their overall model has a relative distance error $\frac{D_m - D_i}{D_i}$ derived from 189~pulsars for which distances are known: $D_m$ is the model distance based on the observed DM and $D_i$ is the independently determined distance (e.g. by parallax). For their sample, the root-mean-square of the relative distance error is 0.398. From version~1.65 of the Australia Telescope National Facility Pulsar Catalogue\cite{2005AJ....129.1993M}, we extracted data for the five pulsars with independently-measured distances within $10^\circ$ of \src{}. For these pulsars, we found that the root-mean-square of the relative distance error is 0.462. Thus we conclude that there is a $\sim$45\,\% error on the estimated distance of 5.7\,kpc. The NE~2001 electron density model\cite{2004ASPC..317..211C} produces a distance estimate of 4.8\,kpc, i.e. indistinguishable within the measured error.

\subsection{Pulsar P-Pdot diagram and Death Valley}

In \Fig~\ref{fig:p_pdot} we report a $P - \dot{P}$ diagram comparing different isolated radio pulsar classes. Grey arrows show upper limits on $\dot{P}$ for those sources having only an upper limit on their measurements. Data to produce these plots were collected from a large number of publications\cite{2005AJ....129.1993M,2006Sci...313..814D,2016ApJ...828L..13R,2016MNRAS.463.2394D, 2014ApJS..212....6O, 2018MNRAS.474..961C,2022Natur.601..526H,2022NatAs...6..828C}.
\noindent
In this manuscript we assume for neutron stars a radius of $R_\mathrm{NS} = 12$\,km radius and $M_\mathrm{NS} = 1.4M_{\odot}$ (compatible with recent measurements\cite{2019Univ....5..159L,2019ApJ...887L..21R}). For all pulsars we derive their surface magnetic field at the pole (see Fig~\ref{fig:p_pdot} color bar) using their $P$ and $\dot{P}$ measurements via the classical dipolar loss formula: 
$$B = \sqrt{\frac{3 c^3 I P \dot{P}}{2 \pi^2 R^6}}.$$ 

Radio emission from pulsars is typically interpreted in terms of pair production in the magnetosphere that begins immediately above the polar caps \cite{1975ApJ...196...51R}. Since the first formulations of this model, it was clear that there should be a limiting period and magnetic field for which a radio pulsar can no longer produce pairs, and radio emission should then cease. The parameter space in the $P - \dot{P}$ diagram where radio emission should start quenching is called "death valley" \cite{1993ApJ...402..264C, 2000ApJ...531L.135Z}. It comprises a large variety of death lines depending on the magnetic field configuration (i.e. dipolar, multi-polar, twisted, etc.), seed $\gamma$-ray photons for pair productions (i.e. curvature or inverse Compton photons), the pulsar obliquity or the assumed stellar radius and moment of inertia. These parameters might well differ from pulsar to pulsar, but all classical formations of death-lines are expected to lie within the so-called "death valley". 
\noindent
In \Fig~\ref{fig:p_pdot} we report the ``death valley'' for pulsars reporting explicitly five of the death lines, three as calculated by \cite{1993ApJ...402..264C} (the numbers 1, 2, and 4 in this work) and two by \cite{2000ApJ...531L.135Z} (labelled by these authors as III and III$^{\prime}$). We assume different parameters: a pure dipole scenario, a dipole with curved field lines, and a twisted magnetic spot at the polar cap in the extreme case where the B field strength at the surface is 10 times stronger than the dipolar field.

\end{methods}


\begin{addendum}
 \item
 We thank the anonymous referees for their comments, which improved the quality of this paper. This scientific work uses data obtained from Inyarrimanha Ilgari Bundara / the Murchison Radio-astronomy Observatory. We acknowledge the Wajarri Yamaji People as the Traditional Owners and native title holders of the Observatory site. The Australian SKA Pathfinder, the Australia Telescope Compact Array, and Parkes/Murriyang are part of the Australia Telescope National Facility (https://ror.org/05qajvd42) which is managed by CSIRO. Operation of ASKAP is funded by the Australian Government with support from the National Collaborative Research Infrastructure Strategy. ASKAP uses the resources of the Pawsey Supercomputing Centre. Establishment of ASKAP, the Murchison Radio-astronomy Observatory and the Pawsey Supercomputing Centre are initiatives of the Australian Government, with support from the Government of Western Australia and the Science and Industry Endowment Fund.
 We acknowledge the Gomeroi people as the Traditional Owners of the ATCA Observatory site, and the Wiradjuri people as the Traditional Owners of the Parkes Observatory site.
Support for the operation of the Murchison Widefield Array is provided by the Australian Government (NCRIS), under a contract to Curtin University administered by Astronomy Australia Limited.
The authors would like to thank SARAO for the approval of the MeerKAT DDT request. The MeerKAT telescope is
operated by the South African Radio Astronomy Observatory, which is a facility of the National Research Foundation, an agency of the Department of Science and Innovation (DSI).
These observations used the FBFUSE and APSUSE computing clusters for data acquisition, storage and analysis. These clusters were funded and installed by the Max-Planck-Institut f\"{u}r Radioastronomie (MPIfR) and the Max-Planck-Gesellschaft.
Breakthrough Listen is managed by the Breakthrough Initiatives, sponsored by the Breakthrough Prize Foundation.
 The Giant Metrewave Radio Telescope is run by the National Centre for Astrophysics of the Tata Institute of Fundamental Research.
 The National Radio Astronomy Observatory is a facility of the National Science Foundation operated under cooperative agreement by Associated Universities, Inc. This project was supported by resources and expertise provided by CSIRO IMT Scientific Computing. Basic research in Radio Astronomy at the U.S.\ Naval Research Laboratory is supported by 6.1 Base Funding. Construction and installation of VLITE was supported by the NRL Sustainment Restoration and Maintenance fund.
 This research is based on observations obtained with \emph{XMM-Newton}, an ESA science mission with instruments and contributions directly funded by ESA Member States and NASA. We thank N.~Schartel for approving our DDT request and the \emph{XMM-Newton} Science Operations Centre for carrying out the observation. 
This research is also based on observations made with the Gran Telescopio Canarias (GTC), installed at the Spanish Observatorio del Roque de los Muchachos of the Instituto de Astrof\'isica de Canarias (IAC) in the island of La Palma, under Director’s Discretionary Time (code GTC03-22BDDT). The EMIR project is led by the IAC with the participation of the Laboratoire d'Astrophysique - Observatoire Midi-Pyrenees (France), Universidad Complutense de Madrid and the Laboratoire d'Astrophysique - Observatoire de Marselle (France). We acknowledge the GTC Director, A. Cabrera, for accepting our DDT request, and thank him, N. Garc\'ia, and T. Mu\~noz-Darias for the useful insights on EMIR data analysis. We would like to thank V. Morello for valuable input on the timing results, A. Harding, R. Turolla and J. Pons for discussion about pulsar magnetospheres, C. Pardo, M. Ronchi, V. Graber and A. Ibrahim for useful discussions on death lines, and M. Sokolowski and C. Trott for commenting on the manuscript as part of the MWA Collaboration review.

 N.H.-W. is the recipient of an Australian Research Council Future Fellowship (project number FT190100231). N.~R. is supported by the European Research Council (ERC) via the Consolidator Grant “MAGNESIA” under grant agreement No. 817661. F.~C.Z. is supported by a Ramon y Cajal Fellowship (grant agreement RYC2021-030888-I). N.~R and F.~C.Z. are partially supported by the program Unidad de Excelencia Mar\'ia de Maeztu CEX2020-001058-M. G.E.A. is the recipient of an Australian Research Council Discovery Early Career Researcher Award (project number DE180100346). D~d.M. acknowledges financial support from the Italian Space Agency (ASI) and National Institute for Astrophysics (INAF). B.W.S. acknowledges funding from the European Research Council (ERC)
under the European Union’s Horizon 2020 research and innovation pro-
gramme (grant agreement No 694745).  KMR acknowledges support from the Vici research programme ``ARGO'' with project number 639.043.815, financed by the Dutch Research Council (NWO).
 \item[Author Contributions] N.H.-W. led the radio proposals, calibrated and processed the MWA and MeerKAT continuum observations, made the initial detection of the source, determined its position and flux density in all radio data, performed the timing error analysis, and prepared the manuscript with contributions from all co-authors. N.R. led the XMM and GTC proposals and preliminary analysis, and contributed theoretical interpretation. S.M. performed the dedispersive and barycentric corrections, and integrated the radio flux density spectrum. B.W.M. performed the timing analysis. E.L. performed the ASKAP data reduction, MWA polarisation analysis, and jointly with I.H., the MeerKAT polarisation analysis. S.D.H. and S.G. performed the GMRT and VLA pre-2016 archival searches. Y.P., K.M.R. and E.B. analysed the MeerKAT PTSUSE and APSUSE data, determining the DM and RM. T.E.C. and S.G. performed the VLITE archive search and imaging. F.C.Z. contributed to the XMM and GTC proposals, performed the XMM analysis and wrote the corresponding text, and with N.R., D.d.M. and M.d.O, analysed the GTC data. D.C.P. and N.D.R.B. conducted the Parkes observing and analysis. C.H. helped develop the filtering and transient source-finding that enabled the detection of the source, with supervision and additional code contributions by N.H.-W., T.J.G., and J.M. G.E.A. and T.J.G. performed the ATCA observations and analysis. A.B. contributed archival X-ray and optical searches. M.C., K.M.R., and B.W.S. assisted with the planning, analysis and interpretation of the high-time resolution MeerKAT data. A.W. performed the MWA observing. 
 \item[Competing Interests] The authors declare that they have no
competing financial interests.
 \item[Data Availability]
 Data that supports this paper are available at the following public repository: \newline\href{https://github.com/nhurleywalker/GPMTransient}{https://github.com/nhurleywalker/GPMTransient}. Further data products can be supplied by the authors on request; raw data can be obtained from the observatory archives through their data portals (see Methods).
  \item[Code Availability] Code that supports this paper is available at the following public repository: \newline\href{https://github.com/nhurleywalker/GPMTransient}{https://github.com/nhurleywalker/GPMTransient}.
   \item[Correspondence] Correspondence and requests for materials should be addressed to N.H.-W. (nhw@icrar.org)

\end{addendum}

\section*{Extended Data}


\renewcommand\thefigure{\arabic{figure}} 
\setcounter{figure}{0}
\renewcommand{\figurename}{Extended Data Figure}

\begin{figure}[H]
\centering
\includegraphics[width=120mm]{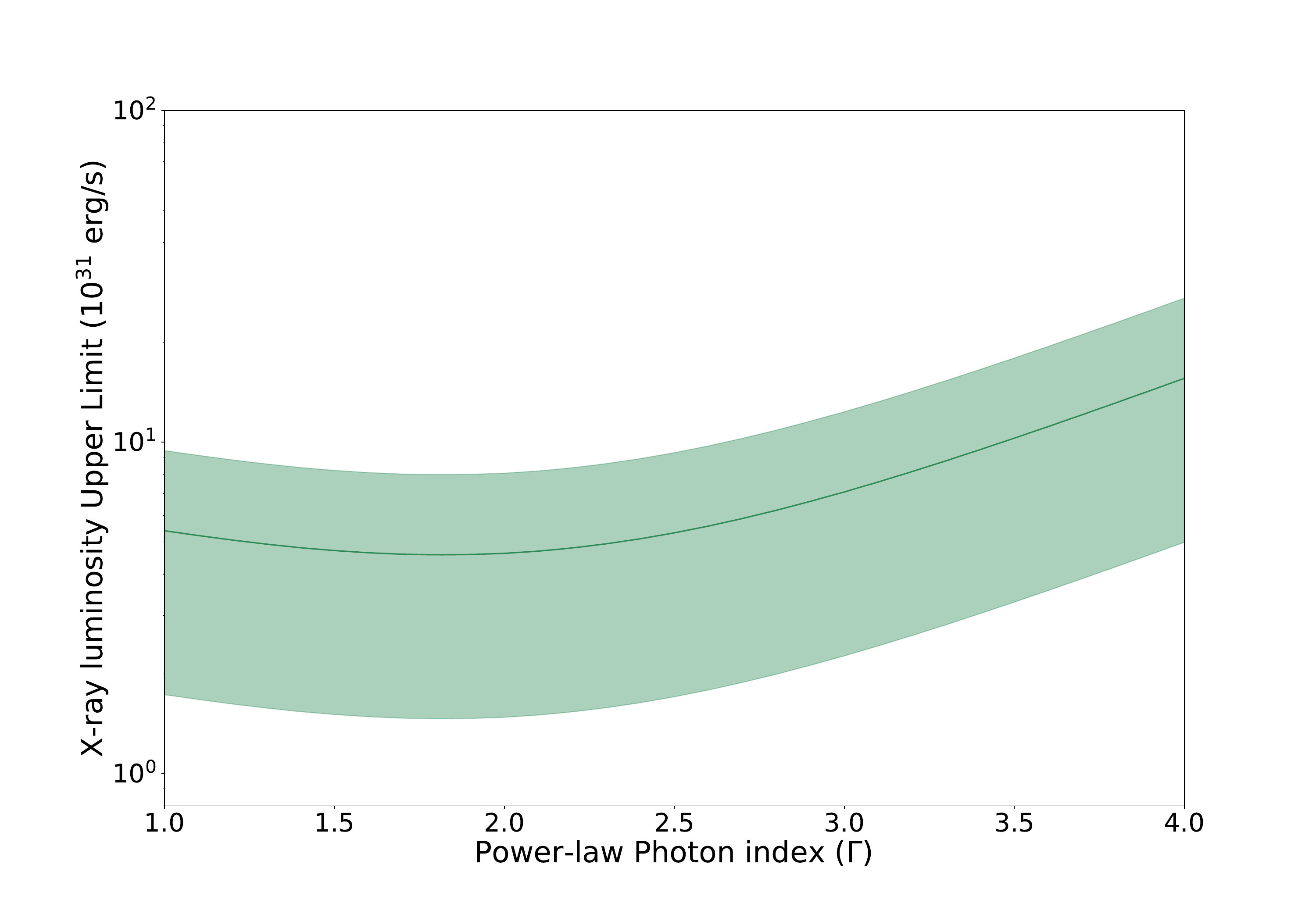}
\includegraphics[width=120mm]{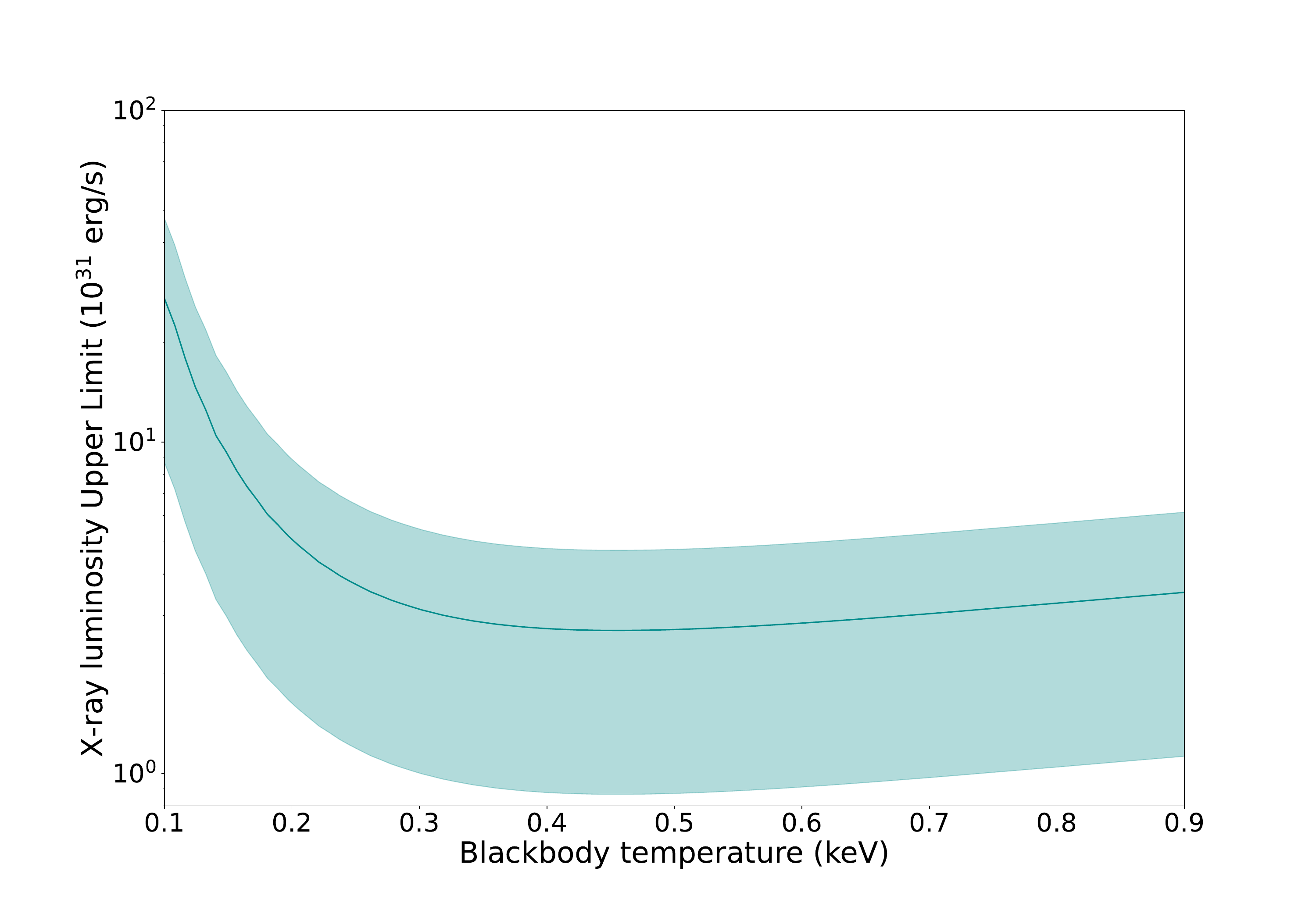}
\caption{3-$\sigma$ upper limits on the persistent X-ray luminosity at the position of \src\ as a function of the assumed spectral shape. The shaded region is for distances in the 3--8\,kpc range. }\label{fig:xray_limits}
\end{figure}

\begin{figure}[H]
\centering
\includegraphics[width=120mm]{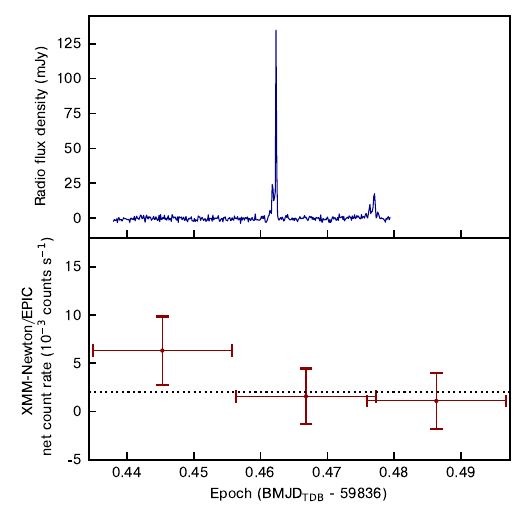}
\caption{Simultaneous radio (ASKAP) and X-ray (XMM-Newton) observations of two bright radio pulses detected on 2022-09-14. The dotted line is the 3$\sigma$ upper limit on the net count rate derived considering the whole observation (see Methods).\label{fig:radioxray_bursts}}
\end{figure}

\begin{figure}[H]
\centering
\includegraphics[width=120mm]{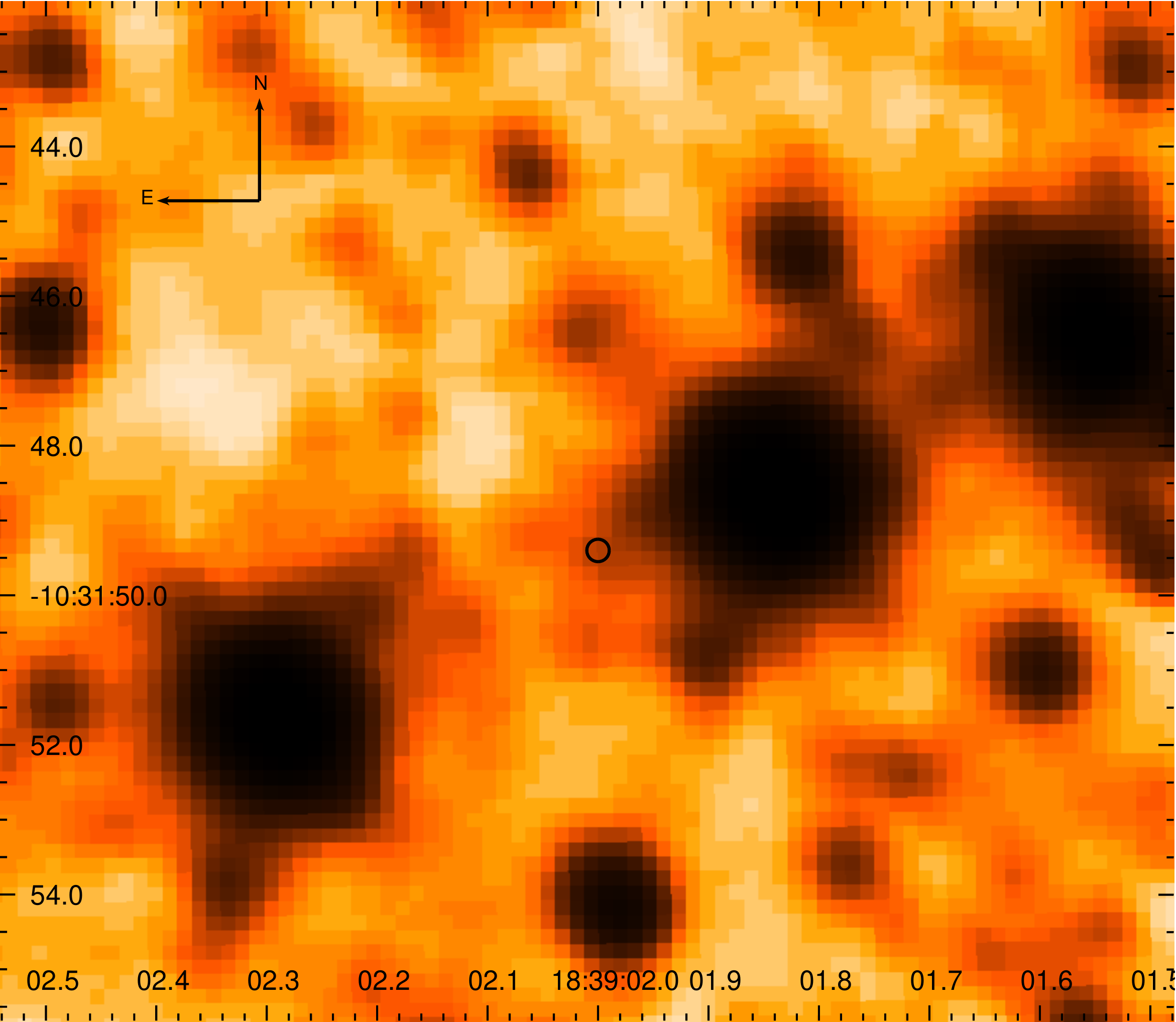}
\caption{Gran Telescopio Canarias EMIR image of the field around \src{} in the $K_s$-band. The black circle represents the source positional error of 0\farcs15.}
\label{fig:NIR_image}
\end{figure}

\begin{figure}[H]
    \centering
    \includegraphics[width=8.9cm]{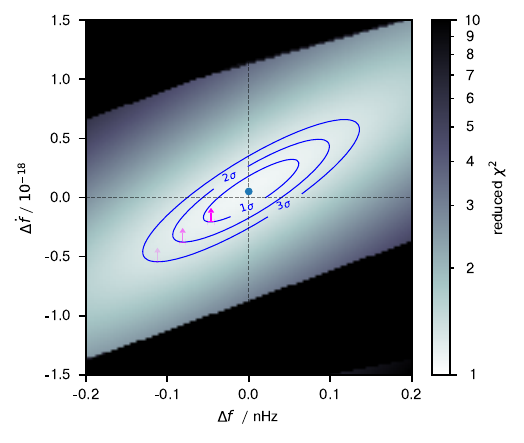}
    \caption{The explored search space in $f$ and $\dot{f}$ for the pulses recorded from \src{}. Contours show the confidence intervals derived from the timing analysis (see Methods). The reduced $\chi^2$ values are shown in greyscale. The central marker shows the best-fit $f = \srcFlong{}$\,Hz and $\dot{f} = \srcFdotlong{}$. The arrow markers show limits on $\dot{f}$; the 1-$\sigma$ limit is shown as an upper limit on $\dot{P}$ in \Fig~\ref{fig:p_pdot}.} 
    \label{fig:ppdot_search}
\end{figure}

\begin{figure}[H]
    \centering
    \includegraphics[width=0.6\textwidth]{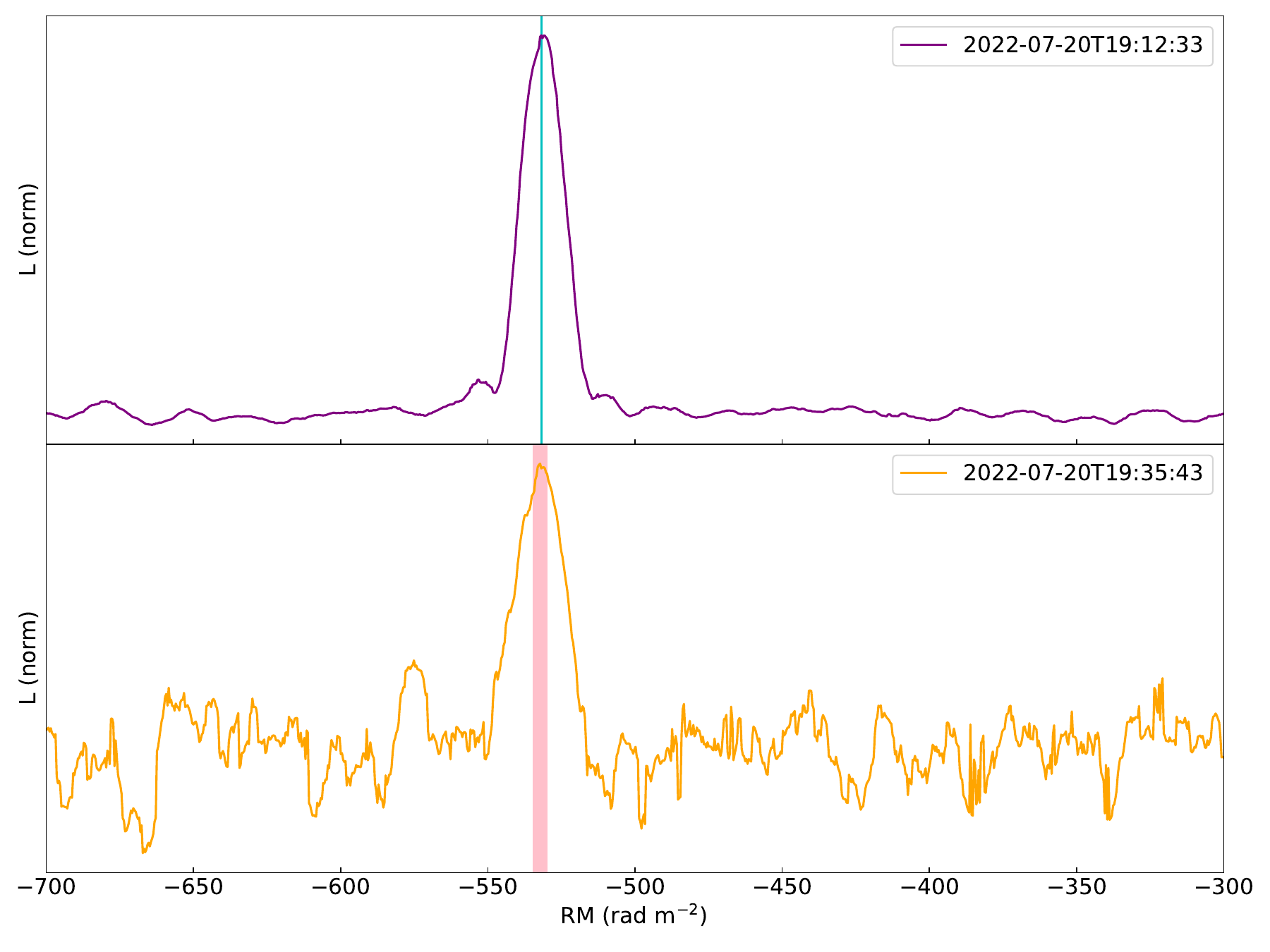}
    \caption{The RM synthesis results of two bursts in the first two pulses detected by MeerKAT, at UTC 2022-07-20T19:12:33 and 2022-07-20T19:35:43. The vertical shaded area represents the measured errors of the RMs. The RM estimations yield from the plot in two periods are $-531.83\pm 0.14$ and $-532.2\pm2.2\ \mathrm{rad\ m}^{-2}$, respectively.} 
    \label{fig:rmfit}
\end{figure}

\begin{figure}[H]
    \centering
    \includegraphics[width=0.6\textwidth]{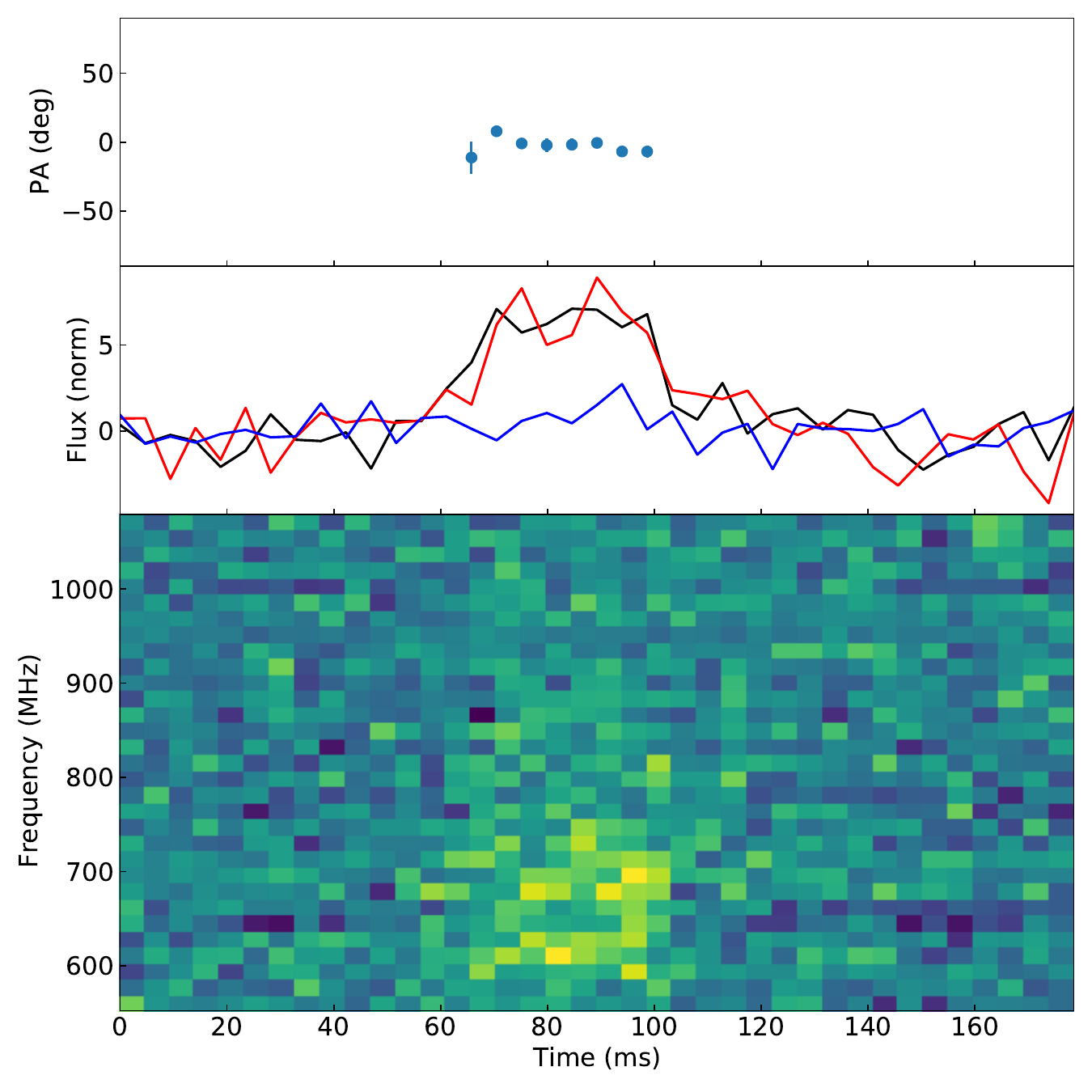}
    \caption{The polarization profile of a 30-ms burst observed with MeerKAT PTUSE. The RM is $-531$\,\RMunit{}. The top panel shows the measurements of the position angle (PA) at different pulse phases. The middle panel shows the total, linearly polarized, and circularly polarized flux densities as a function of time, represented by the black line, the red line and the blue line, respectively. The bottom panel shows the dynamic spectrum of the burst. The start time of the plot is UTC 2022-07-20T19:35:43.228133. The apparent steep spectrum is not intrinsic to the source, but due to the misalignment of the coherent beam.} 
    \label{fig:polarization}
\end{figure}

\begin{figure}[H]
\centering
\includegraphics[width=89mm]{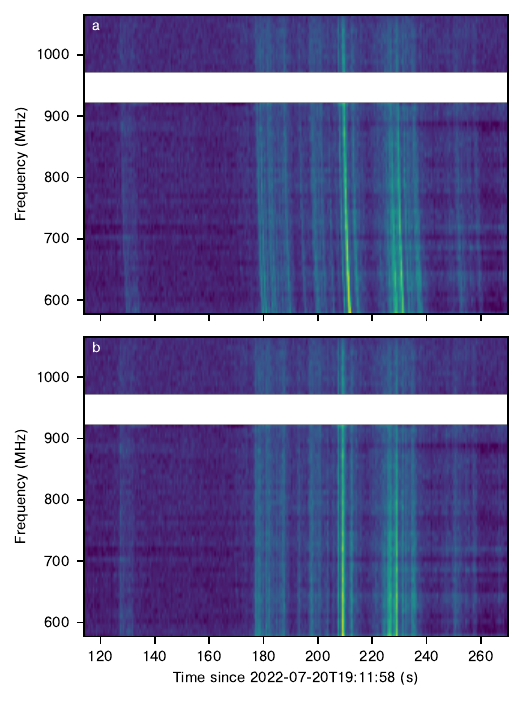}
\caption{Dynamic spectrum of the 2022-07-20T19:12:33 pulse detected with the APSUSE instrument on MeerKAT. The time resolution is 3.9\,ms and the frequency resolution is 8.5\,MHz. Strong interference signals have been removed in the 950-MHz band and at the band edges below 577\,MHz and above 1065\,MHz. The data are shown a) before de-dispersion and b) after de-dispersion.\label{fig:dynspec}}
\end{figure}

\begin{figure}[H]
\centering
\includegraphics[width=178mm]{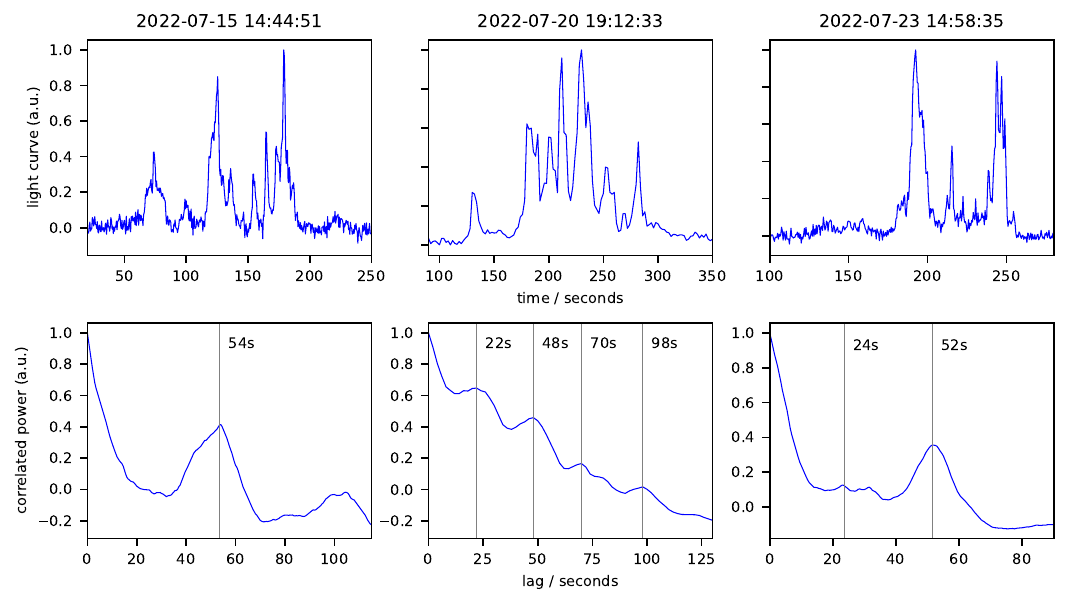}
\caption{Auto-correlation (lag) analysis of three high-signal-to-noise pulses observed with the MWA (left and right panels) and MeerKAT (middle panel). The top row shows the de-dispersed, frequency-integrated light curves as a function of time from the start of each observation, shown in the panel titles. The light curves have been normalised for readability; observational parameters, peak flux densities, and fluences for these pulses are reported in the Supplementary Information Table. The bottom row shows the auto-correlations over the range of timescales well-sampled by each light curve. To indicate the range of observed behaviours, some of the lag peaks are marked with grey vertical lines. \label{fig:acf}}
\end{figure}

\begin{figure}[H]
    \centering
    \includegraphics[width=8.9cm]{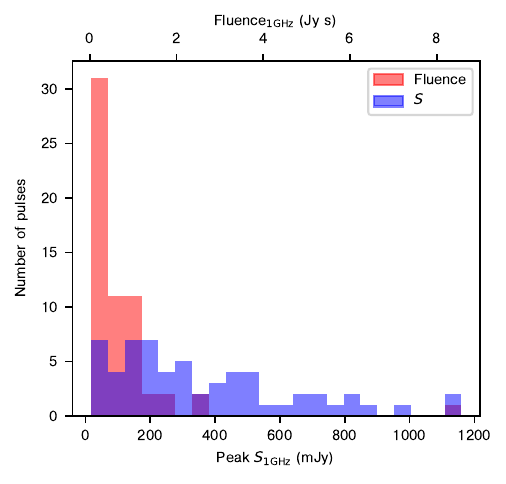}
    \caption{Histogram of flux densities and fluences of detected pulses, using the data provided in the Supplementary Table. The values were scaled to a common frequency of 1\,GHz via \Eqn~\ref{eq:snu} in Methods. 
    \label{fig:flux_hist}}
\end{figure}

\end{document}


\thispagestyle{empty}
\include{preamble}


\begin{table}
\scriptsize
\centering
\caption{\scriptsize \setstretch{0.4} Individual pulse statistics. For pulses split across multiple observations, only a single barycentred time-of-arrival is given, which is used in the timing analysis. Peak flux densities are the maxima observed in a given light curve. Fluences are calculated by multiplying the light curve by the sample time; for faint pulses where the light curves are dominated by noise, only flux densities $>3\sigma$ are used. Flux densities and fluences are also displayed scaled to a common frequency of 1\,GHz using \Eqn~1 in Methods. }
\label{tbl:pulses}
\def\arraystretch{0.6}
\begin{longtable}{c|cccccccc}
Pulse & Observation & Barycentered & Telescope & Frequency & \multicolumn{2}{c}{Peak flux density} & \multicolumn{2}{c}{Fluence} \\
number & start time & TOA & & & At freq & At 1 GHz & At freq & At 1 GHz \\
 & (UTC) & (MJD) & & (MHz) & (Jy) & (mJy) & (Jy s) & (Jy s) \\
\hline
1 & 1988-09-23 22:12 & 47427.92600 & VLA & 327 & 6.43 & 374 & 64.3 & 3.74 \\
301475 & 2001-04-28 11:30 & 52027.48315 & VLA & 327 & 0.974 & 56.7 & 28.0 & 1.63 \\
301480 & 2001-04-28 13:21 & 52027.55989 & VLA & 327 & 1.19 & 69.5 & 47.4 & 2.76 \\
309643 & 2001-08-31 02:20 & 52152.10184 & VLA & 327 & 2.01 & 117 & 106 & 6.18 \\
326337 & 2002-05-12 19:05 & 52406.80046 & GMRT & 240 & 3.42 & 116 & 245 & 8.32 \\
326343 & 2002-05-12 21:16 & 52406.89146 & GMRT & 240 & 0.956 & 32.4 & 107 & 3.63 \\
458391 & 2007-11-17 12:54 & 54421.53419 & GMRT & 240 & 4.43 & 150 & 210 & 7.12 \\
459178 & 2007-11-29 13:06 & 54433.54160 & GMRT & 240 & 1.44 & 49.0 & 24.2 & 0.823 \\
459179 & 2007-11-29 13:27 & 54433.55676 & GMRT & 240 & 1.54 & 52.3 & 114 & 3.87 \\
460158 & 2007-12-14 11:58 & 54448.49375 & GMRT & 240 & 4.73 & 161 & 80.4 & 2.73 \\
596691 & 2013-08-27 13:26 & 56531.56287 & MWA & 118 & 8.9 & 131 & 270 & 3.99 \\
613363 & 2014-05-08 22:05 & 56785.92460 & GMRT & 322 & 1.0 & 56.5 & 42.7 & 2.41 \\
629052 & 2015-01-03 07:06 & 57025.29156 & GMRT & 322 & 2.04 & 115 & 80.7 & 4.56 \\
629059 & 2015-01-03 09:41 & 57025.39809 & GMRT & 322 & 1.84 & 104 & 13.6 & 0.77 \\
629960 & 2015-01-17 03:33 & 57039.14477 & GMRT & 322 & 6.08 & 344 & 107 & 6.06 \\
629967 & 2015-01-17 06:08 & 57039.25046 & GMRT & 322 & 0.744 & 42.1 & 3.1 & 0.175 \\
\multirow{2}{*}{707985} & 2018-04-21 13:27 & \multirow{2}{*}{58229.56424} & VLITE & 341 & \multirow{2}{*}{3.83} & \multirow{2}{*}{242} & \multirow{2}{*}{26.3} & \multirow{2}{*}{1.66} \\
  & 2018-04-21 13:29 & & VLITE & 341 & & & & \\
\multirow{2}{*}{711401} & 2018-06-12 16:11 & \multirow{2}{*}{58281.68106} & MWA & 154 & \multirow{2}{*}{1.84} & \multirow{2}{*}{34.8} & \multirow{2}{*}{101} & \multirow{2}{*}{1.92} \\
  & 2018-06-12 16:13 & & MWA & 185 & & & & \\
776151 & 2021-02-24 13:37 & 59269.56489 & VLITE & 341 & 1.45 & 91.6 & 14.3 & 0.9 \\
803613 & 2022-04-19 13:09 & 59688.55009 & VLITE & 341 & 1.02 & 64.2 & 21.9 & 1.38 \\
804716 & 2022-05-06 08:59 & 59705.37894 & VLITE & 341 & 3.25 & 205 & 47.6 & 3.0 \\
805500 & 2022-05-18 08:03 & 59717.33988 & VLITE & 341 & 0.743 & 46.8 & 9.46 & 0.597 \\
805576 & 2022-05-19 11:53 & 59718.49964 & VLITE & 341 & 2.05 & 129 & 32.6 & 2.05 \\
808340 & 2022-06-30 15:52 & 59760.66994 & MWA & 200 & 3.38 & 87.9 & 45.4 & 1.18 \\
808341 & 2022-06-30 16:17 & 59760.68470 & MWA & 201 & 2.04 & 53.2 & 33.8 & 0.882 \\
809254 & 2022-07-14 14:36 & 59774.61467 & MWA & 200 & 1.42 & 36.9 & 45.2 & 1.18 \\
809255 & 2022-07-14 14:56 & 59774.62898 & MWA & 200 & 1.62 & 42.0 & 49.5 & 1.29 \\
809319 & 2022-07-15 14:25 & 59775.60647 & MWA & 155 & 1.66 & 31.5 & 12.8 & 0.243 \\
809320 & 2022-07-15 14:44 & 59775.62127 & MWA & 185 & 8.67 & 203 & 201 & 4.7 \\
\multirow{2}{*}{809321} & 2022-07-15 15:04 & \multirow{2}{*}{59775.63615} & MWA & 216 & \multirow{2}{*}{3.25} & \multirow{2}{*}{94.2} & \multirow{2}{*}{36.7} & \multirow{2}{*}{1.06} \\
  & 2022-07-15 15:09 & & MWA & 155 & & & & \\
\multirow{2}{*}{809325} & 2022-07-15 16:33 & \multirow{2}{*}{59775.69780} & MWA & 216 & \multirow{2}{*}{2.43} & \multirow{2}{*}{70.4} & \multirow{2}{*}{110} & \multirow{2}{*}{3.2} \\
  & 2022-07-15 16:38 & & MWA & 155 & & & & \\
\multirow{2}{*}{809326} & 2022-07-15 16:53 & \multirow{2}{*}{59775.71247} & MWA & 155 & \multirow{2}{*}{3.23} & \multirow{2}{*}{61.3} & \multirow{2}{*}{63.5} & \multirow{2}{*}{1.2} \\
  & 2022-07-15 16:58 & & MWA & 185 & & & & \\
\multirow{2}{*}{809445} & 2022-07-17 12:28 & \multirow{2}{*}{59777.52809} & MWA & 216 & \multirow{2}{*}{3.63} & \multirow{2}{*}{105} & \multirow{2}{*}{110} & \multirow{2}{*}{3.17} \\
  & 2022-07-17 12:33 & & MWA & 155 & & & & \\
809445 & 2022-07-17 12:31 & 59777.52770 & Murriyang & 1792 & 0.15 & 1160 & 0.193 & 1.48 \\
809446 & 2022-07-17 12:53 & 59777.54308 & MWA & 185 & 2.52 & 59.2 & 40.5 & 0.95 \\
809660 & 2022-07-20 19:12 & 59780.80837 & MeerKAT & 807 & 0.106 & 54.9 & 6.32 & 3.29 \\
809778 & 2022-07-22 14:27 & 59782.60850 & MWA & 154 & 1.19 & 22.4 & 4.01 & 0.0757 \\
\multirow{2}{*}{809779} & 2022-07-22 14:47 & \multirow{2}{*}{59782.62433} & MWA & 118 & \multirow{2}{*}{10.3} & \multirow{2}{*}{153} & \multirow{2}{*}{236} & \multirow{2}{*}{3.49} \\
  & 2022-07-22 14:52 & & MWA & 154 & & & & \\
809780 & 2022-07-22 15:12 & 59782.63932 & MWA & 119 & 2.81 & 41.6 & 118 & 1.75 \\
\multirow{2}{*}{809845} & 2022-07-23 14:58 & \multirow{2}{*}{59783.63146} & MWA & 216 & \multirow{2}{*}{8.02} & \multirow{2}{*}{232} & \multirow{2}{*}{197} & \multirow{2}{*}{5.7} \\
  & 2022-07-23 15:03 & & MWA & 88 & & & & \\
\multirow{2}{*}{809846} & 2022-07-23 15:18 & \multirow{2}{*}{59783.64589} & MWA & 185 & \multirow{2}{*}{1.96} & \multirow{2}{*}{46.0} & \multirow{2}{*}{64.8} & \multirow{2}{*}{1.52} \\
  & 2022-07-23 15:23 & & MWA & 216 & & & & \\
809972 & 2022-07-25 13:26 & 59785.56789 & MWA & 154 & 1.35 & 25.5 & 5.09 & 0.0963 \\
810233 & 2022-07-29 13:06 & 59789.55111 & MWA & 118 & 8.69 & 128 & 362 & 5.35 \\
810234 & 2022-07-29 13:25 & 59789.56591 & MWA & 88 & 5.06 & 62.3 & 414 & 5.1 \\
\multirow{2}{*}{810299} & 2022-07-30 13:12 & \multirow{2}{*}{59790.55783} & MWA & 185 & \multirow{2}{*}{6.1} & \multirow{2}{*}{143} & \multirow{2}{*}{156} & \multirow{2}{*}{3.64} \\
  & 2022-07-30 13:16 & & MWA & 216 & & & & \\
\multirow{2}{*}{810300} & 2022-07-30 13:31 & \multirow{2}{*}{59790.57236} & MWA & 154 & \multirow{2}{*}{3.27} & \multirow{2}{*}{61.8} & \multirow{2}{*}{107} & \multirow{2}{*}{2.02} \\
  & 2022-07-30 13:36 & & MWA & 185 & & & & \\
810304 & 2022-07-30 15:04 & 59790.63353 & MWA & 185 & 2.18 & 51.1 & 52.9 & 1.24 \\
\multirow{2}{*}{810305} & 2022-07-30 15:24 & \multirow{2}{*}{59790.64929} & MWA & 185 & \multirow{2}{*}{2.83} & \multirow{2}{*}{66.1} & \multirow{2}{*}{88.5} & \multirow{2}{*}{2.07} \\
  & 2022-07-30 15:29 & & MWA & 185 & & & & \\
\multirow{2}{*}{810306} & 2022-07-30 15:44 & \multirow{2}{*}{59790.66415} & MWA & 185 & \multirow{2}{*}{1.6} & \multirow{2}{*}{37.3} & \multirow{2}{*}{4.94} & \multirow{2}{*}{0.116} \\
  & 2022-07-30 15:49 & & MWA & 185 & & & & \\
810495 & 2022-08-02 13:00 & 59793.54919 & MWA & 185 & 1.24 & 29.1 & 53.6 & 1.25 \\
\multirow{2}{*}{810496} & 2022-08-02 13:20 & \multirow{2}{*}{59793.56316} & MWA & 154 & \multirow{2}{*}{4.99} & \multirow{2}{*}{94.5} & \multirow{2}{*}{165} & \multirow{2}{*}{3.13} \\
  & 2022-08-02 13:24 & & MWA & 185 & & & & \\
810497 & 2022-08-02 13:44 & 59793.57818 & MWA & 154 & 1.79 & 33.9 & 40.8 & 0.771 \\
810561 & 2022-08-03 13:11 & 59794.55567 & MWA & 118 & 1.46 & 21.5 & 2.0 & 0.0296 \\
\multirow{2}{*}{810562} & 2022-08-03 13:30 & \multirow{2}{*}{59794.57023} & MWA & 88 & \multirow{2}{*}{11.2} & \multirow{2}{*}{137} & \multirow{2}{*}{404} & \multirow{2}{*}{4.97} \\
  & 2022-08-03 13:35 & & MWA & 118 & & & & \\
\multirow{2}{*}{810563} & 2022-08-03 13:50 & \multirow{2}{*}{59794.58458} & MWA & 216 & \multirow{2}{*}{5.06} & \multirow{2}{*}{146} & \multirow{2}{*}{224} & \multirow{2}{*}{6.49} \\
  & 2022-08-03 13:55 & & MWA & 88 & & & & \\
810566 & 2022-08-03 14:59 & 59794.63070 & MWA & 185 & 0.746 & 17.5 & 23.8 & 0.558 \\
\multirow{2}{*}{810567} & 2022-08-03 15:19 & \multirow{2}{*}{59794.64712} & MWA & 154 & \multirow{2}{*}{2.33} & \multirow{2}{*}{44.1} & \multirow{2}{*}{118} & \multirow{2}{*}{2.23} \\
  & 2022-08-03 15:24 & & MWA & 185 & & & & \\
810756 & 2022-08-06 12:34 & 59797.53000 & MWA & 118 & 2.21 & 32.6 & 3.5 & 0.0517 \\
\multirow{2}{*}{810758} & 2022-08-06 13:19 & \multirow{2}{*}{59797.56101} & MWA & 88 & \multirow{2}{*}{9.99} & \multirow{2}{*}{123} & \multirow{2}{*}{176} & \multirow{2}{*}{2.16} \\
  & 2022-08-06 13:24 & & MWA & 118 & & & & \\
\multirow{2}{*}{810759} & 2022-08-06 13:38 & \multirow{2}{*}{59797.57564} & MWA & 216 & \multirow{2}{*}{4.91} & \multirow{2}{*}{142} & \multirow{2}{*}{297} & \multirow{2}{*}{8.57} \\
  & 2022-08-06 13:43 & & MWA & 88 & & & & \\
\end{longtable}
\end{table}

%% file: preamble.tex
\newcommand{\fig}{Fig.}
\newcommand{\figs}{Figs.}
\newcommand{\Fig}{Fig.}
\newcommand{\EFig}{Extended Data Fig.}
\newcommand{\Figs}{Figs.}
\newcommand{\sect}{Section}
\newcommand{\sects}{Sections}
\newcommand{\Sect}{Section}
\newcommand{\Sects}{Sections}
\newcommand{\tab}{Table}
\newcommand{\tabs}{Tables}
\newcommand{\Tab}{Table}
\newcommand{\Tabs}{Tables}
\newcommand{\eqn}{equation}
\newcommand{\eqns}{equations}
\newcommand{\Eqn}{Equation}
\newcommand{\Eqns}{Equations}
\newcommand{\etal}{et~al.}

\newcommand{\farcm}{\mbox{\ensuremath{.\mkern-4mu^\prime}}}
\newcommand{\farcs}{\mbox{\ensuremath{.\!\!^{\prime\prime}}}}
\newcommand{\fdg}{\mbox{\ensuremath{.\!\!^\circ}}}

\newcommand{\src}{GPM\,J\,1839\ensuremath{-}10}
\newcommand{\GLX}{GLEAM-X\,J\,162759.5\ensuremath{-}523504.3}
\newcommand{\PSR}{PSR\,J\,0901\ensuremath{-}4046}
\newcommand{\srcDM}{\ensuremath{273.5}}
\newcommand{\DMerror}{\ensuremath{2.6}}
\newcommand{\DMunit}{pc\,cm\ensuremath{^{-3}}}
\newcommand{\srcDist}{\ensuremath{5.7}}
\newcommand{\Disterror}{\ensuremath{2.9}}
\newcommand{\Distunit}{kpc}
\newcommand{\srcRM}{\ensuremath{-573}}
\newcommand{\RMerror}{\ensuremath{1}}
\newcommand{\RMunit}{rad\,m\ensuremath{^{-2}}}
\newcommand{\srcPlong}{\ensuremath{1318.1957}}
\newcommand{\srcPerr}{\ensuremath{0.0002}}
\newcommand{\srcP}{\ensuremath{1318}}
\newcommand{\srcFlong}{\ensuremath{0.000758612(7)}}
\newcommand{\srcFerr}{\ensuremath{1.2\times10^{-10}}}
\newcommand{\srcFdotlong}{\ensuremath{5\times10^{-20}}}

\newcommand{\srcalpha}{\ensuremath{-3.17}}
\newcommand{\srcalphaerr}{\ensuremath{0.06}}
\newcommand{\srcq}{\ensuremath{-0.56}}
\newcommand{\srcqerr}{\ensuremath{0.03}}

\newcommand{\ndetections}{71}
\newcommand{\srcPdot}{\ensuremath{3.6\times10^{-13}}}
\newcommand{\srcPdotts}{\ensuremath{9.5\times10^{-13}}}
\newcommand{\Pdotunit}{\,s\,s\ensuremath{^{-1}}}

\newcommand{\srcBp}
{\ensuremath{1.2\times10^{15}}}
\newcommand{\srcBpts}
{\ensuremath{2.4\times10^{15}}}

\newcommand{\srcBpwd}{\ensuremath{4\times10^{9}}}
\newcommand{\srcBptswd}{\ensuremath{8.1\times10^{9}}}

\newcommand{\srcSLum}{\ensuremath{1.3\times10^{25}}}
\newcommand{\srcSLumts}{\ensuremath{6\times10^{25}}}

\newcommand{\srcSLumwd}{\ensuremath{2.4\times10^{30}}}
\newcommand{\srcSLumtswd}{\ensuremath{1\times10^{31}}}

\newcommand{\srctau}{\ensuremath{58}}
\newcommand{\srctauts}{\ensuremath{22}}

\newcommand{\srcRLum}{\ensuremath{10^{28}}}

\newcommand{\ergpers}{erg\,s\ensuremath{^{-1}}}
\newcommand{\flux}{erg\,cm\ensuremath{^{-2}}\,s\ensuremath{^{-1}}}

\def\arc{\mbox{$^{\prime\prime}$}}
\def\nh{\hbox{$N_{\rm H}$}}